\documentclass[ALICE,manyauthors]{cernphprep}

\usepackage{graphicx}
\usepackage{tabularx}
\usepackage{booktabs}
\usepackage{footnote}
\usepackage{threeparttable}  
\usepackage{multirow}
\usepackage{bm}
\usepackage{relsize}
\usepackage{gensymb}

\usepackage{chngcntr}

\usepackage{amsmath}
\usepackage{amssymb}

\newcommand*\rot{\rotatebox{90}}
\newcommand{\ml}[1]{\ensuremath{#1}}

\usepackage{array}
\newcolumntype{L}[1]{>{\raggedrightlet\newline\\arraybackslashhspace{0pt}}m{#1}}
\newcolumntype{C}[1]{>{\centering\arraybackslash}p{#1}}
\newcolumntype{R}[1]{>{\raggedleftlet\newline\\arraybackslashhspace{0pt}}m{#1}}
\newcolumntype{Y}{>{\raggedleft\arraybackslash}X}

\usepackage[comma,square,numbers,sort&compress]{natbib}
\usepackage{hyperref}
\usepackage{lineno}

\newcommand{\Rppb}{\ensuremath{R_{\rm pPb}}\xspace}

\newcommand{\snnt}[1]{\ensuremath{\sqrt{s_{\rm NN}} = #1 \text{\,TeV}}\xspace}

\newcommand{\sppt}[1]{\ensuremath{\sqrt{s} = #1 \text{\,TeV}}\xspace}

\newcommand{\gevc}[1]{\ensuremath{#1 \text{\,GeV/$c$}}\xspace}

\newcommand{\pp}{pp\xspace}
\newcommand{\ppb}{p--Pb\xspace}
\newcommand{\pbpb}{Pb--Pb\xspace}

\newcommand{\pt}{\ensuremath{p_{\text{T}}}\xspace}

\newcommand{\dedx}{\ensuremath{\text{d}E/\text{d}x}\xspace}
\newcommand{\mdedx}{\ensuremath{\left <\text{d}E/\text{d}x \right>}\xspace}

\newcommand{\res}{\ensuremath{\sigma_{\text{d}E/\text{d}x}}\xspace}

\newcommand{\chpi}{\ensuremath{\pi^{+}+\pi^{-}}\xspace}
\newcommand{\chk}{\ensuremath{{\rm K}^{+}+{\rm K}^{-}}\xspace}
\newcommand{\chp}{\ensuremath{{\rm p}+{\rm \bar{p}}}\xspace}

\begin{document}%


\begin{titlepage}
\PHyear{2016}
\PHnumber{003}      
\PHdate{05 January}  
%

\title{Multiplicity dependence of charged pion, kaon, and (anti)proton production
  at large transverse momentum \\ in p--Pb collisions at  $\mathbf{\sqrt{{\textit s}_{\rm NN}}}$=5.02 TeV}
  \ShortTitle{Pion, kaon, and (anti)proton production in p--Pb collisions} 

\Collaboration{ALICE Collaboration\thanks{See Appendix~\ref{app:collab} for the list of collaboration members}}
\ShortAuthor{ALICE Collaboration} 


\begin{abstract}

The production of charged pions, kaons and (anti)protons has been measured at mid-rapidity ($-0.5 < y < 0$) in p--Pb collisions at  $\sqrt{s_{\rm NN}}\,=5.02$\,TeV using the ALICE detector at the LHC. Exploiting particle identification capabilities at high transverse momentum ($p_{\rm T}$), the previously published $p_{\rm T}$ spectra have been extended to include measurements up to 20\,GeV/$c$ for seven event multiplicity classes.  The $p_{\rm T}$ spectra for pp collisions at $\sqrt{s}=7$\,TeV, needed to interpolate a pp reference spectrum, have also been extended up to 20\,GeV/$c$ to measure the nuclear modification factor ($R_{\rm pPb}$) in non-single diffractive p--Pb collisions. \\At intermediate transverse momentum ($2 < p_{\rm T} < 10$\,GeV/$c$) the
proton-to-pion ratio increases with multiplicity in p--Pb collisions, a similar effect is not present in the kaon-to-pion ratio. The $p_{\rm T}$ dependent structure of such increase is qualitatively similar to those observed in pp and heavy-ion collisions.  At high $p_{\rm T}$  ($>10$\,GeV/$c$), the particle ratios  are consistent with those reported for pp and Pb--Pb collisions at the LHC energies. \\ At intermediate $p_{\rm T}$ the (anti)proton $R_{\rm pPb}$ shows a Cronin-like enhancement, while pions and kaons show little or no nuclear modification. At high $p_{\rm T}$ the charged pion, kaon and (anti)proton  $R_{\rm pPb}$ are consistent with unity within statistical and systematic uncertainties.

\end{abstract}
\end{titlepage}
\setcounter{page}{2}

%
%
%
%


\section{Introduction}

In heavy-ion collisions at ultra-relativistic energies, it is well established
that a strongly coupled Quark-Gluon-Plasma (sQGP) is formed~\cite{Adams:2005dq, Adcox:2004mh, Arsene:2004fa, Back:2004je, Schukraft:2011na}.  Some of the characteristic features of the sQGP are strong collective flow and opacity to jets. The collective behavior is observed both as an azimuthal anisotropy of produced particles~\cite{ALICE:2011ab}, where the magnitude is described by almost ideal (reversible) hydrodynamics, and as a hardening of \pt
spectra for heavier hadrons, such as protons, by radial flow~\cite{Abelev:2012wca}. Jet quenching is observed as a reduction of both high \pt particles~\cite{Abelev:2012hxa,CMS:2012aa} and also fully reconstructed jets~\cite{Abelev:2013kqa}. The
interpretation of these sQGP properties requires comparisons with
reference measurements like \pp and p-A collisions. Recent measurements
in high multiplicity \pp, p-A and d-A collisions at different energies
have revealed strong flow-like effects even in these small systems~\cite{Abelev:2012sk,Chatrchyan:2013eya,Khachatryan:2010gv,Chatrchyan2013795,Abelev:2012ola,Aad:2012gla,Aad:2013fja,Chatrchyan:2013nka,PhysRevC.88.024906,Adams:2006nd}. The origin of these phenomena is debated ~\cite{Shuryak:2013ke, Werner:2013ipa,Bozek:2013ska, Dumitru:2010iy, Schenke:2015aqa, Ma:2014pva,Sjostrand:2007gs, Corke:2010yf, PhysRevLett.111.042001} and the data reported here provide further inputs to this
discussion. 

In a previous work, we reported the evidence of radial flow-like patterns in \ppb collisions~\cite{Abelev:2013haa}. This effect was found to increase with increasing event multiplicity and to be qualitatively consistent with calculations which incorporate the hydrodynamical evolution of the system. It was also discussed that in small systems, mechanisms like color-reconnection may produce radial flow-like effects. The present paper
reports complementary measurements covering the intermediate \pt region (2--10\,GeV/$c$) and the high-\pt region (10--20\,GeV/$c$) exploiting the capabilities of the High Momentum
Particle Identification Detector (HMPID) and the Time Projection Chamber
(TPC). In this way,  high precision measurements are achieved in the intermediate  \pt region where cold nuclear matter effects like the Cronin
enhancement~\cite{Kopeliovich:2002yh,Hwa:2004zd} have been reported by previous
experiments~\cite{PhysRevLett.31.1426,PhysRevD.11.3105}, and where the
particle ratios, e.g., the proton (kaon) production relative to that
of pions, are affected by large final state effects in central \pbpb collisions~\cite{Abelev:2014laa}.
Particle ratios are expected to be modified by flow, but hydrodynamics
is typically expected to be applicable only up to a few GeV/$c$~\cite{Huovinen:2001cy}. At higher \pt, ideas such as parton recombination have been
proposed leading to baryon-meson effects~\cite{Das:1977cp}. In this way the new dataset complements the lower \pt results.

In addition, particle identification at large transverse momenta in \ppb collisions provides new constraints on the nuclear parton distribution functions (nPDF) which are key inputs in interpreting a large amount of experimental data like d-Au and deep inelastic scattering~\cite{Armesto:2015lrg}. Finally, the measurement is also important to study the particle species dependency of the nuclear modification factor (\Rppb), to better understand parton energy loss mechanisms in heavy-ion collisions.

In this paper, the charged pion, kaon and (anti)proton \Rppb are reported for non-single diffractive (NSD) \ppb collisions. The \pp reference spectra for this measurement were obtained using interpolations of data at different collision energies. The  already published \pt spectra for inelastic (INEL) \pp collisions at \sppt{7}~\cite{Adam:2015qaa} were extended up to \gevc{20} and the results are presented here for the first time. These measurements together with the results for INEL \pp collisions at \sppt{2.76} ($\pt<20$\,GeV/$c$)~\cite{Abelev:2014laa} were used to determine \pp reference spectra at \sppt{5.02} using the interpolation method described in~\cite{Abelev:2013ala}.


The paper is organized as follows. In Sec.~\ref{sec:data} the ALICE detector as well as the event and track selections are discussed. The analysis procedures for particle identification using the HMPID and TPC detectors are outlined in Sec.~\ref{sec:hmpid} and Sec.~\ref{sec:tpc}, respectively. Section~\ref{sec:results} presents the results and
discussions. Finally, Sec.~\ref{sec:conclusions} summarizes the main results.


\section{Data sample, event and track selection}
\label{sec:data}

The results are obtained using data collected with the ALICE detector during
the 2013 \ppb run at \snnt{5.02}. The detailed description of the ALICE
detector can be found in~\cite{1748-0221-3-08-S08002} and the performance during run 1 (2009--2013) is described in~\cite{Abelev:2014ffa}. Because
of the LHC 2-in-1 magnet design, it is impossible to adjust the energy of the proton and lead-ion beams independently. They are $4$\,TeV per $Z$ which gives different
energies due to the different $Z/A$ of the colliding protons and lead ions. The
nucleon-nucleon center-of-mass system is moving in the laboratory frame with a rapidity of $y_{\rm NN}$ = -0.465 in the direction of the proton beam rapidity. In the following, $y_{\rm lab}$ ($\eta_{\rm lab}$) are used to indicate the (pseudo)rapidity in the laboratory reference frame, whereas $y$ ($\eta$) denotes the (pseudo)rapidity in the center-of-mass reference system where the Pb beam is assigned positive rapidity.

In the analysis of the \ppb data, the event selection follows that used in the analysis of inclusive charged particle production~\cite{ALICE:2012mj}. The minimum bias (MB) trigger signal was provided by
the V0 counters~\cite{1748-0221-8-10-P10016}, which contain two arrays of 32
scintillator tiles each covering the full azimuth within $2.8 < \eta_{\rm lab}
< 5.1$ (V0A) and $-3.7 < \eta_{\rm lab} < -1.7$ (V0C). The signal amplitude and arrival time collected in each
tile were recorded. A coincidence of signals in both V0A and V0C detectors was
required to remove contamination from single diffractive and electromagnetic
events. In the offline analysis, background events were further suppressed by requiring the arrival time of signals on the neutron Zero Degree Calorimeter A, which is positioned in the Pb-going direction, to be compatible with a nominal \ppb collision occurring close to the nominal interaction point. The estimated mean number of interactions per bunch crossing was below 1\% in the sample chosen for this analysis. Due to the weak correlation between collision geometry and multiplicity, the
particle production in \ppb collisions is studied in event multiplicity classes instead of centralities~\cite{Adam:2014qja}. The multiplicity classes are defined
using the total charge deposited in the V0A detector as in~\cite{Abelev:2013haa}, where V0A is positioned in the Pb-going direction.  The MB results have been normalized to the total number of NSD events using a trigger and vertex reconstruction efficiency correction which amounts to $3.6\,\%\pm3.1\,\%$~\cite{ALICE:2012xs}. The multiplicity dependent results have been normalized to the visible (triggered) cross-section correcting for the vertex reconstruction efficiency (this was not done in~\cite{Abelev:2013haa}). This correction is of the order of  5\% for the lowest V0A multiplicity class (80--100\%) and negligible for the other multiplicity classes ($<1\%$). 


In the \sppt{7} \pp analysis the MB trigger required a hit in the
two innermost layers of the Inner Tracking System (ITS), the Silicon Pixel
Detector (SPD), or in at least one of the V0 scintillator arrays in
coincidence with the arrival of proton bunches from both directions. The
offline analysis to eliminate background was done using the time information
provided by the V0 detectors in correlation with the number of clusters and
tracklets\footnote{Tracklets are pairs of hits from the two layers of the SPD which make a line
pointing back to the collision vertex.} in the SPD.

Tracks are required to be reconstructed in both the ITS and the TPC. Additional track selection criteria are the same as in~\cite{Adam:2015kca} and based on
the number of space points, the quality of the track fit, and the distance of
closest approach to the reconstructed collision vertex.  Charged tracks where the identity of the particle has changed due to a
weak decay, e.g., ${\rm K}^{-} \rightarrow \mu^{-} + \bar{\nu}_{\mu}$, are identified by the tracking algorithm due to their distinct kink
topologies~\cite{Aamodt:2011zj} and rejected in this analysis. The remaining contamination is negligible ($\ll$1\%). In order to have the
same kinematic coverage as used in the \ppb low \pt analysis~\cite{Abelev:2013haa}, the
tracks were selected in the pseudorapidity interval $-0.5 < \eta < 0$. In addition, for the HMPID analysis it is required that the tracks are
propagated and matched to a primary ionization cluster in the Multi-Wire
Proportional Chamber (MWPC) gap of the HMPID detector \cite{Adam:2015qaa,
  Adam:2015kca}. 

The published results of charged pion, kaon and (anti)proton production at low \pt for \pp~\cite{Adam:2015qaa}
and \ppb~\cite{Abelev:2013haa} collisions at \sppt{7} and \snnt{5.02}, respectively, used different
Particle IDentification (PID) detectors and techniques. A summary of the \pt ranges covered by the
published analyses and the analyses presented in this paper can be found in
Table~\ref{tab:1}.

\begin{table}[htb]
  \let\center\empty
  \let\endcenter\relax
  \centering
  \resizebox{0.6\textwidth}{!}{
  \begin{threeparttable}
  \begin{tabular}{c l c c c c c c c}
  \toprule
  \midrule [1.5pt]
  & \textbf{Analysis} & & $\pi^{+}+\pi^{-}$ & & ${\rm K}^{+}+{\rm K}^{-}$ & & ${\rm p}+\bar{\rm p}$ & \\
  \cmidrule{2-2} \cmidrule{4-4} \cmidrule{6-6} \cmidrule{8-8}\\[-5pt]
					 & Published~\cite{Adam:2015qaa}\tnote{*}		& & $0.1-3.0$	& & $0.2-6.0$	& & $0.3-6.0$ 	&	\\ [2pt]
  \rot{\rlap{\textbf{pp}}} 	 & TPC  ${\rm d}E$/${\rm d}x$  	rel. rise & & $2-20$ 	& & $3-20$	& & $3-20$ 	&	\\ [2pt]
					 & 				& & 		& & 		& & 		&  	\\
					 & Published~\cite{Abelev:2013haa}\tnote{$\dagger$} 	& & $0.1-3.0$	& & $0.2-2.5$	& & $0.3-4.0$ 	&	\\ [2pt]
					 & HMPID 			& & $1.5-4.0$	& & $1.5-4.0$	& & $1.5-6.0$ 	&	\\ [2pt]
  \rot{\rlap{\textbf{p--Pb}}} 	 &  TPC  ${\rm d}E$/${\rm d}x$  	rel. rise & & $2-20$	& & $3-20$	& & $3-20$ 	&	\\ [2pt]
  \midrule [1.5pt]
  \bottomrule
  \end{tabular}
  \begin{tablenotes}
  \item[*] Included detectors: ITS, TPC, Time-of-Flight (TOF), HMPID. The results also include the kink-topology identification of the weak decays of charged kaons.
  \item[$\dagger$] Included detectors: ITS, TPC, TOF.
  \end{tablenotes}
  \end{threeparttable}
  \label{table:PID}
}
  \caption{Transverse momentum ranges (GeV/$c$) covered by the individual and combined analyses for pp collisions at $\sqrt{s}\,=7$\,TeV and p--Pb collisions at $\sqrt{s_{\rm NN}}\,=5.02$\,TeV.}
\label{tab:1}
\end{table}

In the following, the analysis techniques used to obtain the identified particle \pt spectra in the intermediate and high-\pt ranges using HMPID and TPC will be discussed.


\section{HMPID analysis}
\label{sec:hmpid}

\begin{figure}[t!p]
  \centering
  \includegraphics[width=0.75\textwidth]{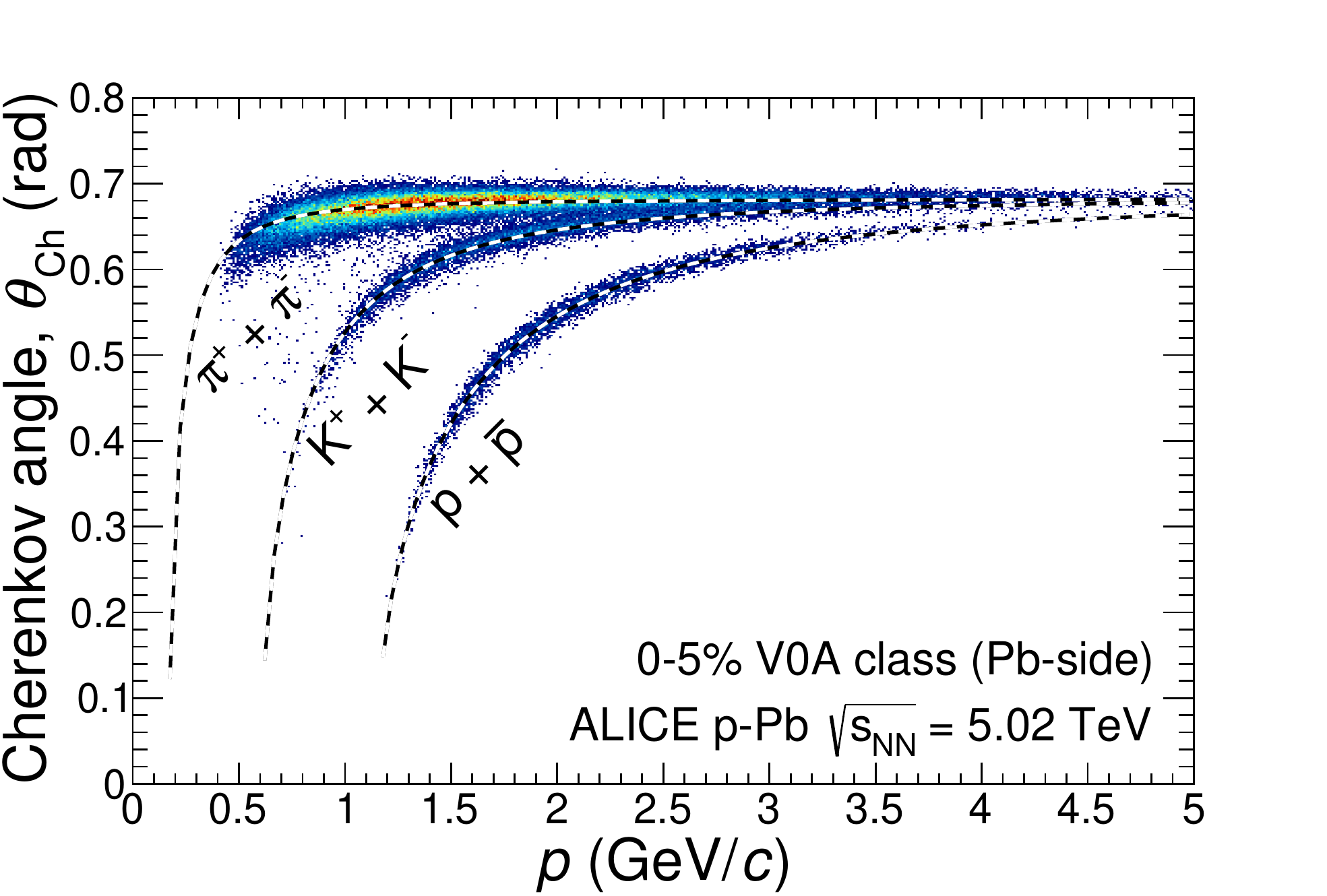}
  \caption{(Color online.) Cherenkov angle measured in the HMPID as a function of the track momentum in p--Pb collisions at  $\sqrt{s_{\rm NN}}=\,5$.02\,TeV
  for the 0--5\% V0A multiplicity class (see the text for further details). The dashed lines represent the expected curves calculated using Eq.~\ref{costheta} for each particle species.}
  \label{AnglevsMom}
\end{figure}

\begin{figure}[t!p]
  \centering
  \includegraphics[width=0.8\textwidth]{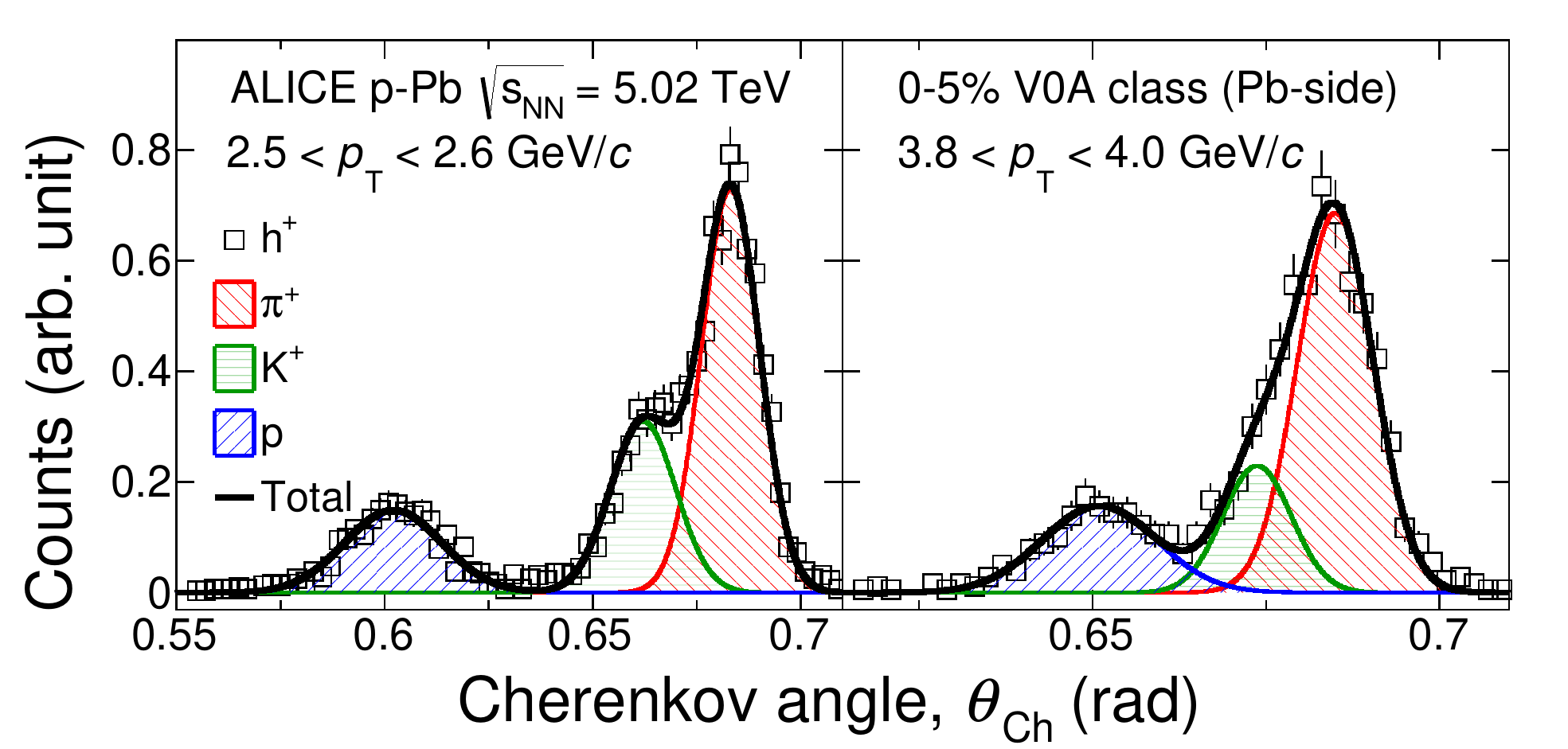}
  \caption{(Color online.) Distributions of the Cherenkov angle measured in the HMPID for positive tracks having $p_{\rm{T}}$ between 2.5--2.6\,GeV/$c$ (left) and between 3.8--4.0\,GeV/$c$ (right), 
 in p--Pb collisions at  $\sqrt{s_{\rm NN}}=5$.02\,TeV for the 0--5\% V0A multiplicity class (see the text for further details).}
  \label{Fits}
\end{figure}

The HMPID detector~\cite{Piuz:381431} is located about 5 m from the beam axis, covering a limited acceptance of $|\eta_{\rm{lab}}|<0.5$ and 1.2$^{\circ}$~$<$ $\phi$ $<$~58.5$^{\circ}$, that corresponds to $\sim$5\% of the TPC geometrical acceptance ($2\pi$ in azimuthal angle and the pseudo-rapidity interval $|\eta|<0.9$~\cite{Alme2010316}) for high $p_{\rm{T}}$  tracks. The HMPID analysis uses $\sim$9$\times$10$^{7}$ minimum-bias \ppb events at \snnt{5.02}.   The event and track selection  and the analysis technique are similar to those described in~\cite{Adam:2015qaa, Adam:2015kca}. It is required that tracks are propagated and matched to a primary ionization cluster in the Multi-Wire Proportional Chamber (MWPC) gap of the HMPID detector. The PID in the HMPID is done by measuring the Cherenkov angle, $\theta_{\rm{Ch}}$ \cite{Piuz:381431}:

\begin{equation}
\cos\theta_{\rm{Ch}} = \frac{1}{n\beta} \Longrightarrow \theta_{\rm{Ch}} = \arccos\Bigg(\frac{\sqrt{p^{2} + m^{2}}}{np} \Bigg),
\label{costheta}
\end{equation}
where $n$ is the refractive index of the radiator used (liquid C$_{6}$F$_{14}$ with $n$ = 1.29 at $E_{\rm ph}$ = 6.68\,eV and temperature $T=20$~\celsius
), $p$ and $m$ are the momentum and the mass of the given particle, respectively. The measurement of the single photon $\theta_{\rm{Ch}}$ angle in the HMPID requires the knowledge of the track parameters, which are estimated by the track extrapolation from the central tracking detectors up to the radiator volume, where the Cherenkov photons are emitted. 
Only one charged particle cluster is associated to each extrapolated track, selected as the closest cluster to the extrapolated track point on the cathode plane. To reject the fake cluster-match associations in the detector, a selection on the distance $d_{(\rm{track-MIP})}$ computed on the cathode plane between the track extrapolation point and the reconstructed charged-particle cluster position is applied. The distance has to be less than 5 cm, independent of track momentum.  Starting from the photon cluster coordinates on the photocathode, a back-tracking algorithm calculates the corresponding emission angle. The Cherenkov photons are selected by the Hough Transform Method  (HTM) \cite{DiBari:2003wy} that discriminates the signal from the background. For a given track, the Cherenkov angle $\theta_{\rm{Ch}}$ is then computed 
as the weighted mean of the single photon angles selected by the HTM.  Figure~\ref{AnglevsMom} shows the $\theta_{\rm{Ch}}$ as a function of the track momentum.  The reconstructed angle distribution for a given momentum interval is fitted by a sum of three Gaussian distributions, corresponding to the signals from pions, kaons, and protons. The fitting is done in two steps. In the first step the initial values of fit parameters are set to the expected
values.  The mean values, $\langle \theta_{\rm{Ch}} \rangle_{i}$, are obtained from Eq. \ref{costheta}, tuning the refractive index to match
the observed Cherenkov angles, and the resolution values are taken from a Monte Carlo simulation of the detector response. After this first step, the $p_{\rm{T}}$ dependences of the mean and width are fitted with the function given by Eq. \ref{costheta}  and a polynomial one, respectively. In the second step, the fitting is repeated with the yields as the only free parameters, constraining the mean and resolution values to the fitted value. The second iteration is particularly important at high $p_{\rm{T}}$ where the separation between different species is reduced. Figure \ref{Fits} gives examples of  fits to the reconstructed Cherenkov angle distributions in two narrow 
$p_{\rm{T}}$ intervals for the 0--5\% multiplicity class.  The raw yields are then corrected  by the total
reconstruction efficiency given by the convolution of the tracking, PID efficiency, and distance cut correction. The tracking efficiency, convoluted with the geometrical acceptance of the detector, has
been evaluated using Monte Carlo simulations. For all three particle species this efficiency increases from $\sim$5\% at 1.5\,GeV/$c$ up to $\sim$6\% at high \pt. The PID efficiency is determined by the Cherenkov angle 
reconstruction efficiency. It has been computed by means of Monte-Carlo simulations and it reaches $\sim$90\% for particles with velocity $\beta$$\sim$1, with no significant difference between positive and negative tracks. The 
distance cut correction, defined as the ratio between the number of the tracks that pass the cut on $d_{(\rm{track-MIP})}$ and all the tracks in the detector acceptance, has been evaluated from data. It is momentum dependent, and it is equal to $\sim$53\% at 1.5 GeV/$\it{c}$, reaches $\sim$70\% for particles with velocity $\beta$$\sim$1. A small difference between positive and 
negative tracks is present; negative tracks having a distance correction $\sim$2\% lower than the positive ones. This effect is caused by a radial residual misalignment of the HMPID chambers and an imperfect estimation of the 
energy loss in the material traversed by the track. Tracking efficiency, PID efficiency and distance cut correction do not show variation with the event track multiplicity.

\subsection{Systematic uncertainties}
The systematic uncertainty on the results of the HMPID analysis has contributions from tracking, PID  and tracks association \cite{Adam:2015qaa, Adam:2015kca}. The uncertainties related to the tracking have been estimated by changing the track selection cuts individually, e.g. the number of crossed readout rows in the TPC and the value of the track's $\chi^{2}$ normalized to the number of TPC clusters. To estimate the PID contribution, the parameters (mean and resolution) of the fit function used to extract the raw yields were varied by a reasonable quantity, leaving them free in a given range; the range chosen for the mean values is [$\langle \theta_{\rm Ch}\rangle-\sigma$, $\langle\theta_{\rm Ch}\rangle+\sigma$] and for the resolution [$\sigma - 0.1\sigma$, $\sigma+0.1\sigma$]. A variation of 10\% of the resolution corresponds to its maximum expected variation when taking into account the different running conditions of the detector during data taking which have an impact on its performance. When the means are changed, the resolution values are fixed to the default value and vice versa. The variation of parameters is done for the three Gaussians (corresponding to the three particle species) simultaneously. In addition, the uncertainty of the association of the track to the charged particle signal 
is obtained by varying the default value of the distance cut required for the match by $\pm$ 1 cm. 
These contributions do not vary with the collision multiplicity. A summary of the different contributions to the systematic uncertainty for the HMPID \ppb analysis is shown in Table~\ref{tab:systematics}.

\begin{table*}
\centering
\begin{tabular}{lcccccc}
\hline
\hline
 Effect         &  \multicolumn{2}{c}{$\pi^{+}+\pi^{-}$}  &   \multicolumn{2}{c}{${\rm K}^{+}+{\rm K}^{-}$}  &  \multicolumn{2}{c}{${\rm p}+\bar{\rm p}$}   \\
\hline
 $p_{\rm{T}}$ value (GeV/$c$) & 2.5  & 4  & 2.5 & 4 & 2.5 & 4 \\
PID  & 6\% & 12\% & 6\% & 12\% & 4\% & 5\% \\
Tracking efficiency  & \multicolumn{2}{c}{6\%} & \multicolumn{2}{c}{6\%} & \multicolumn{2}{c}{7\%} \\
Distance cut correction  & 6\% & 2\% & 6\% & 2\% & 4\% & 2\% \\ 
\hline
\hline
\end{tabular}
\caption{Main sources of systematic uncertainties for the HMPID p--Pb analysis.}
\label{tab:systematics}
\end{table*}

\section{TPC \dedx relativistic rise analysis}\label{subsec:1}
\label{sec:tpc}

The relativistic rise regime of the specific energy loss, \dedx, measured by the
TPC allows identification of charged pions, kaons, and (anti)protons up to
$\pt = \gevc{20}$.  The results presented in this paper were obtained using
the method detailed in~\cite{Adam:2015kca}. In this analysis, around $8\times 10^{7}$
($4.7\times 10^{7}$) \ppb (\pp) MB triggered events were used. The event and track selection has already been discussed in Section~\ref{sec:data}.

\begin{figure}[t!f]
  \centering
  \includegraphics[width=0.75\textwidth]{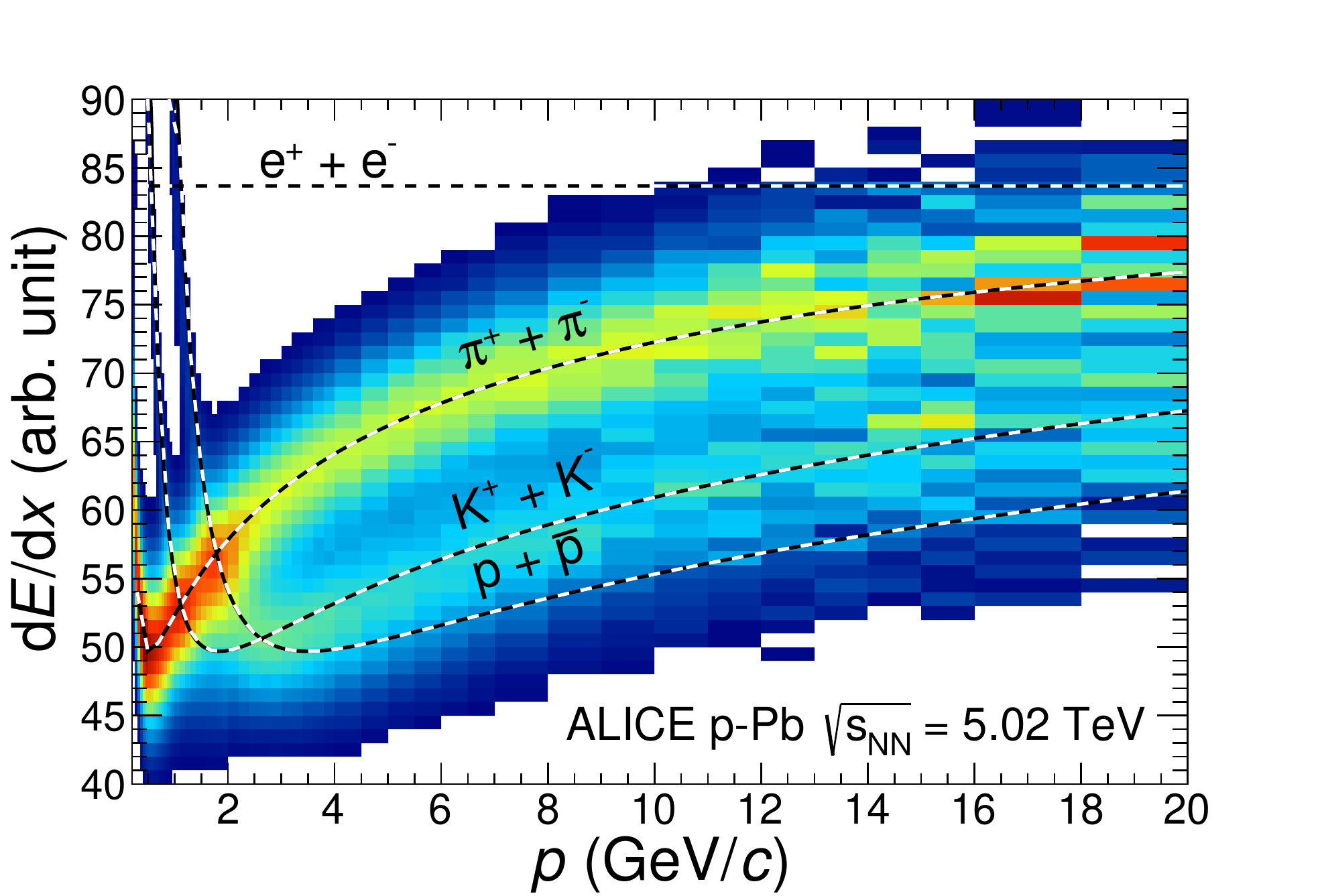}
  \caption{(Color online.) Specific energy loss, ${\ensuremath{\text{d}E/\text{d}x}\xspace}$, as a function of momentum $p$ in the pseudorapidity range
    $-0.5 < \eta < -0.375$ for minimum bias p--Pb collisions. In each momentum bin the ${\ensuremath{\text{d}E/\text{d}x}\xspace}$ spectra have
    been normalized to have unit integrals and only bins with more than 2\% of
    the counts are shown (making electrons not visible in the figure, except
    at very low momentum). The curves show the $\langle {\ensuremath{\text{d}E/\text{d}x}\xspace} \rangle$ response for pions, kaons, protons and electrons.}
  \label{dEdx2DFits}
\end{figure}

\begin{figure}[h!f]
  \centering
  \includegraphics[width=\textwidth]{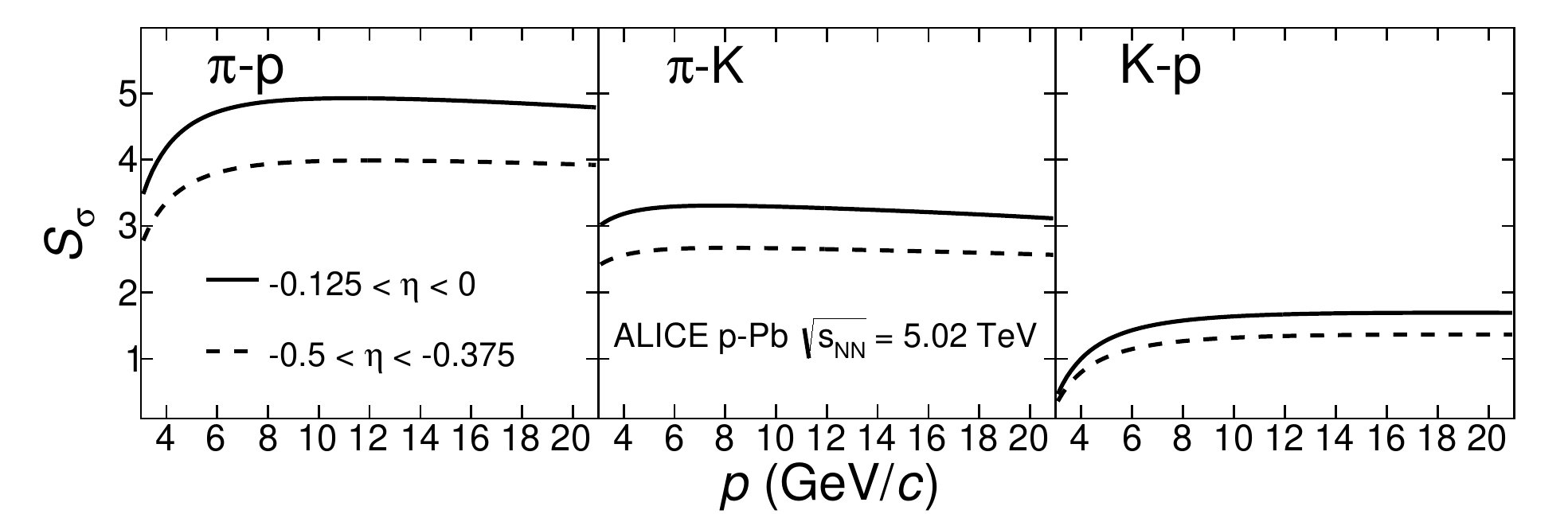}
  \caption{Separation in number of standard deviations between: pions and protons (left panel), pions and kaons (middle panel), and kaons and protons (right panel). Results for minimum bias p--Pb data and for two specific pseudorapidity intervals are shown. More details can be found in~\cite{Adam:2015kca}.}
  \label{nSigmaPPB}
\end{figure}

As discussed in~\cite{Adam:2015kca}, the \dedx is calibrated taking into
account chamber gain variations, track curvature and diffusion, to obtain a
response that essentially only depends on $\beta\gamma$. Inherently, tracks at
forward rapidity will have better resolution due to longer integrated
track-lengths, so in order to analyze homogeneous samples the analysis is performed in
four $\eta$ intervals.  Samples of topologically identified pions (from
${\rm K}^0_{\rm S}$ decays), protons (from $\Lambda$ decays) and electrons (from $\gamma$ conversions)
were used to parametrize the Bethe-Bloch response, \mdedx$(\beta\gamma)$, and
the relative resolution, \res (\mdedx)~\cite{Adam:2015kca}. For the \ppb data, these response functions are found to be slightly multiplicity dependent (the \mdedx changes by $\sim$0.4\% and the sigma by $\sim$2.0\%). However, a single set of functions is used for all multiplicity
intervals, and the dependence is included in the systematic uncertainties. Figure~\ref{dEdx2DFits} shows \dedx as a function of momentum for \ppb events. The characteristic separation power between particle species in number of standard deviations ($S_{\sigma}$) as a function of $p$, is shown in
Fig.~\ref{nSigmaPPB} for  minimum bias \ppb collisions. For example, $S_\sigma$ for pions and kaons is calculated as:

\begin{equation}
\label{eq:rtpc:3}
S_\sigma = \frac{ \left\langle \frac{{\rm d}E}{{\rm d}x} \right\rangle_{\chpi} - \left\langle \frac{{\rm d}E}{{\rm d}x} \right\rangle_{\chk} }{0.5\left( \sigma_{\chpi} + \sigma_{\chk} \right) }.
\end{equation}

The separation in number of standard deviations is the largest (smallest) between pions and
protons (kaons and protons) and it is nearly constant at large momenta.

\begin{figure}[t!f]
  \centering
  \includegraphics[keepaspectratio, width=0.99\columnwidth]{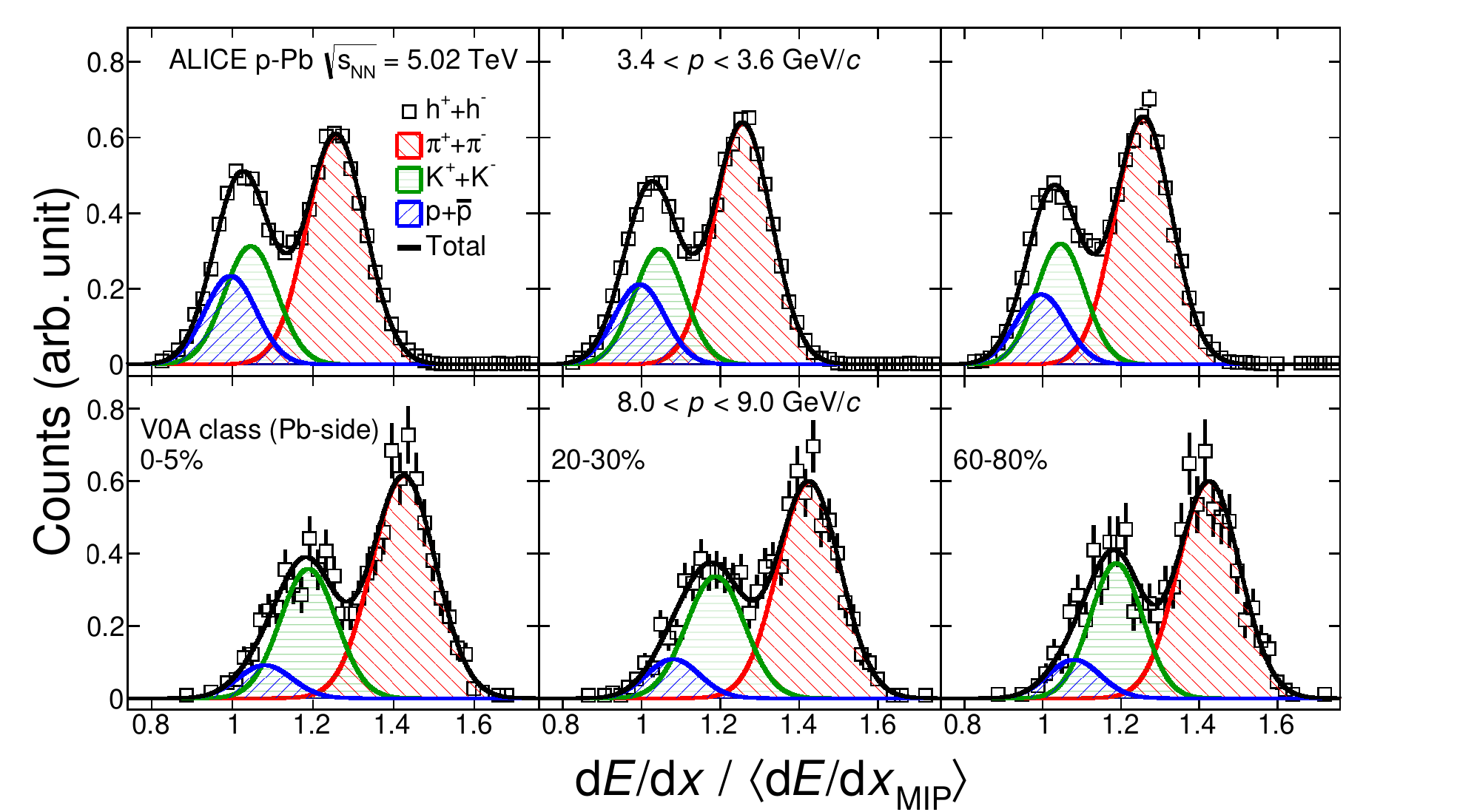}
  \caption{(Color online.) Four-Gaussian fits (lines) to the ${\ensuremath{\text{d}E/\text{d}x}\xspace}$ spectra (markers) for tracks having momentum $3.4 < p < 3.6$\,GeV/$c$ (top row) and $8.0 < p < 9.0$\,GeV/$c$ (bottom row) within $-0.125 < \eta < 0$. All of the spectra are normalized to have unit integrals. Columns refer to different V0A multiplicity classes. Individual signals of charged pions, kaons, and (anti)protons are shown as red, green, and blue dashed areas, respectively. The contribution of electrons is not visible and is negligible ($< 1\%$). }
  \label{fig1:sec1}
\end{figure}

The main part of this analysis is the determination of the relative particle
abundances, hereafter called particle fractions, which are defined as the \chpi, \chk, \chp and $e^{+}+e^{-}$ yields normalized to that for inclusive
charged particles. Since the TPC \dedx signal is Gaussian distributed as illustrated in~\cite{Adam:2015kca}, particle fractions are obtained using four-Gaussian fits to
\dedx distributions in $\eta$ and $p$ intervals. The
parameters (mean and width) of the fits are fixed using the parametrized
Bethe-Bloch and resolution curves mentioned earlier. Examples of these fits can be seen in Fig.~\ref{fig1:sec1} for two momentum intervals,
$3.4 < p < \gevc{3.6}$ and $8 < p < \gevc{9}$. The particle fractions in a \pt range, are obtained as the weighted average of the contributing $p$ intervals. Since the particle fractions as a function of \pt are found to be independent of $\eta$, they are averaged. The particle fractions
measured in \ppb and \pp collisions are corrected for relative efficiency
differences using DPMJET~\cite{Roesler:2000he} and PHOJET~\cite{Engel:1994vs} Monte Carlo (MC) generators,
respectively. Furthermore, the relative pion and proton abundances were corrected for the
contamination of secondary particles (feed-down), more details of the method can be found in~\cite{Adam:2015kca}.

The invariant yields, $1/(2\pi\pt)$ ${\rm d^{2}}N/{\rm d}y{\rm d}\pt$, are constructed using two components: the corrected particle fractions and the corrected invariant charged particle yields. For the \pp analysis at \sppt{7}, the latter component  was taken directly from the published results for inclusive charged particles~\cite{Abelev:2013ala}. However, analogous results for \ppb data are neither available for neither the kinematic range $-0.5 < y < 0$ nor for the different multiplicity classes~\cite{Abelev:2014dsa}, they were therefore measured here and the results used to determine the invariant yields. 

\subsection{Systematic uncertainties}\label{subsubsec:3}

\begin{table}[htb]
  \let\center\empty
  \let\endcenter\relax
  \centering
  \resizebox{\textwidth}{!}{
  \begin{threeparttable}
  \begin{tabular}{c l c@{\hspace{-1.05em}} c@{\hspace{-1.05em}} c c c@{\hspace{-0.5em}} c@{\hspace{-0.5em}} c c c@{\hspace{-0.5em}} c@{\hspace{-0.5em}} c c c@{\hspace{-1.2em}} c@{\hspace{-1.2em}} c c c@{\hspace{-0.53em}} c@{\hspace{-0.53em}} c}
  \toprule
  \midrule [1.5pt]
  & \multicolumn{20}{c}{\textbf{pp collisions}} \\
  \midrule
  &  & \multicolumn{3}{c}{$\ml{\pi}^{+}+\ml{\pi}^{-}$} & & \multicolumn{3}{c}{K$^{+}+$K$^{-}$} & & \multicolumn{3}{c}{${\rm p}+\bar{\rm p}$} & & \multicolumn{3}{c}{ K/$\pi$ } & & \multicolumn{3}{c}{ ${\rm p}/\pi$  }\\
  \cmidrule{3-5} \cmidrule{7-9} \cmidrule{11-13} \cmidrule{15-17} \cmidrule{19-21}\\ [-10pt]
  & {$p_{\rm T}$ (GeV/$c$)} 		& 2.0 &  & 10 &  & 3.0  &  & 10 &  & 3.0 &  & 10 &  & 3.0 &  & 10 &  & 3.0 &  & 10\\ [1pt]
  \midrule \\[-8pt]
  & \textbf{Uncertainty} \\ [5pt]
  & {Event and track selection\tnote{*}} 	 & 	&7.3\% 	& 	&	& 	&7.3\%	& 	&	& 	&7.3\%	& 	&	& 	&$-$	& 	&	& 	&$-$	& 		\\ [2pt]
  & {Feed-down correction} 		 & 	&0.2\%	& 		&	&	&$-$	& 	&	& 	&1.2\%	& 	&	& 	&0.2\%	& 	&	& 	&1.2\%	& 	\\ [2pt]
  & {Efficiency correction} 		& 		&3.2\%	& 		&	& 	&3.2\%	& 	&	& 	&3.2\%	& 	&	& 	&4.5\%	& 	&	& 	&4.5\%	& 	\\ [2pt]
  & {Correction for muons}  		& 0.3\%		&	& 0.5\%		&	& 	&$-$	& 	&	& 	&$-$	& 	&	& 0.3\%	&	& 0.5\%	&	& 0.3\%	&	& 0.5\% 	\\ [2pt]
  & {Parametrization of Bethe-Bloch} 	&     	&  &    	&  	&      		&   	&    	&  &     	&  	&    	&  &  		&  		&    	&  	&     	& 	&   				\\ [2pt]
  & {and resolution curves} 			& 1.8\%		&	& 1.9\%		&	& 20\%	&	& 6.9\%	&	& 24\%	&	& 15\%	&	& 17\%	&	& 9.0\%	&	& 17\%	&	& 19\% 	\\ [2pt]
  \midrule [1.5pt]

  \midrule 
  & \multicolumn{20}{c}{\textbf{p--Pb collisions}} \\
  \midrule
  & \textbf{Uncertainty} \\ [5pt]
  & {Event and track selection\tnote{*}} 	 & 3.3\% 	&	& 3.6\% 	&	& 3.3\%	&	& 3.6\%	&	& 3.3\%	&	& 3.6\%	&	& 	&$-$	& 	&	& 	&$-$	& 		\\ [2pt]
  & {Feed-down correction} 		 & 	&$\leq$~0.2\%	& 		&	&	&$-$	&	&	&2.6\%	&	&0.7\%	&	& 	&$\leq$~0.2\%	&	&	&2.6\%	&	&0.7\% 	\\ [2pt]
  & {Efficiency correction} 		& 		&3.2\%	& 		&	& 	&3.2\%	& 	&	& 	&3.2\%	&	& 	& 	&4.5\%	& 		&	&&4.5\%	&	\\ [2pt]
  & {Correction for muons\tnote{$\dagger$}} 		& 0.3\%		&	& 0.4\%		&	& 	&$-$	&	&	& 	&$-$	&	&	& 0.3\%	&	& 0.4\%	&	& 0.3\%	&	& 0.4\% 	\\ [2pt]
  \midrule
  & {Parametrization of Bethe-Bloch} 	&     	&  &    	&  	&      		&   	&    	&  &     	&  	&    	&  &  		&  		&    	&  	&     	& 	&   				\\ [2pt]
  & {and resolution curves} 	\\ [2pt]
  & \multicolumn{1}{r}{\textbf{   0-5 \%}} 			& 1.7\%		&	& 1.9\%		&	& 17\%	&	& 8.0\%	&	& 15\%	&	& 13\%	&	& 16\%	&	& 10.4\%	&	& 12\%	&	& 11\% 	\\ [2pt]
  & \multicolumn{1}{r}{\textbf{  5-10 \%}} 			& 1.7\%		&	& 2.0\%		&	& 17\%	&	& 5.6\%	&	& 16\%	&	& 12\%	&	& 18\%	&	& 7.2\%	&	& 14\%	&	& 24\% 	\\ [2pt]
  & \multicolumn{1}{r}{\textbf{ 10-20 \%}} 			& 1.6\%		&	& 1.9\%		&	& 16\%	&	& 7.2\%	&	& 16\%	&	& 12\%	&	& 18\%	&	& 9.5\%	&	& 16\%	&	& 15\% 	\\ [2pt]
  & \multicolumn{1}{r}{\textbf{ 20-40 \%}} 			& 1.6\%		&	& 2.0\%		&	& 16\%	&	& 6.7\%	&	& 17\%	&	& 15\%	&	& 18\%	&	& 8.0\%	&	& 17\%	&	& 18\% 	\\ [2pt]
  & \multicolumn{1}{r}{\textbf{ 40-60 \%}} 			& 1.5\%		&	& 1.9\%		&	& 15\%	&	& 6.5\%	&	& 17\%	&	& 12\%	&	& 18\%	&	& 8.3\%	&	& 18\%	&	& 13\% 	\\ [2pt]
  & \multicolumn{1}{r}{\textbf{ 60-80 \%}} 			& 1.6\%		&	& 1.8\%		&	& 16\%	&	& 6.3\%	&	& 20\%	&	& 13\%	&	& 21\%	&	& 8.3\%	&	& 22\%	&	& 18\% 	\\ [2pt]
  \rot{\rlap{Multiplicity classes}} & \multicolumn{1}{r}{\textbf{80-100 \%}} 			& 1.4\%		&	& 1.5\%		&	& 13\%	&	& 5.9\%	&	& 20\%	&	& 13\%	&	& 16\%	&	& 7.3\%	&	& 23\%	&	& 21\% 	\\ [2pt]
  \midrule [1.5pt]
  \bottomrule
  \end{tabular}
  \begin{tablenotes}
  \item[*] Common to all species, values taken from~\cite{Abelev:2014dsa,Abelev:2013ala}.
  \item[$\dagger$] Found to be multiplicity independent.
  \end{tablenotes}
  \end{threeparttable}
  \label{table:systpPb}
}
  \caption{Summary of the systematic uncertainties for the charged pion, kaon, and (anti)proton spectra and for the particle ratios.  Note that ${\rm K/\pi}=(\rm{K}^{+}+\rm{K}^{-})/(\pi^{+}+\pi^{-})$ and ${\rm p}/\pi=(\rm{p}+\bar{\rm{p}})/(\pi^{+}+\pi^{-})$.}
\label{tab:2}  
\end{table}

The systematic uncertainties mainly consist of two components: the first is due to the event and track selection, and the second one is due to the PID. The first component was obtained from the analysis of inclusive charged particles~\cite{Abelev:2013ala,Abelev:2014dsa}. For INEL \pp collisions at 7 TeV, the systematic uncertainties have been taken from~\cite{Abelev:2013ala}. For \ppb collisions, there are no measurements in the $\eta$ interval reported here ($-0.5<\eta<0$); however, it has been shown that the systematic uncertainty exhibits a negligible dependence on $\eta$ and multiplicity~\cite{Adam:2014qja}. Therefore, the systematic uncertainties reported  in~\cite{Abelev:2014dsa} have been assigned to the identified charged hadron \pt spectra for all the V0A multiplicity classes.\\
The second component was measured following the procedure described in~\cite{Adam:2015kca}, where the largest contribution is attributed to the uncertainties in the parameterization of the Bethe-Bloch and resolution curves used to constrain the fits. The uncertainty is calculated by varying the \mdedx and \res in the particle fraction fits (Fig.~\ref{fig1:sec1}) within the precision of the \dedx response calibration, $\sim$1\% and 5\% for \mdedx and \res, respectively. A small fraction of this uncertainty was found to be multiplicity dependent, it was estimated as done in the previous ALICE publication~\cite{Abelev:2013haa}.

A summary of the main systematic uncertainties on the \pt spectra and the particle ratios for \ppb and \pp collisions can be found in Table~\ref{tab:2} for two \pt intervals. For pions, the main contribution is related to event and track selection and the associated common corrections. In the case of kaons and protons the largest uncertainty is attributed to the parametrization of the \dedx response. For kaons, the uncertainty decreases with \pt and increases with multiplicity while for protons the multiplicity dependence is opposite. This variation mainly reflects the changes in the particle ratios with \pt and multiplicity. 


\begin{figure}[th!f]
  \centering
  \includegraphics[keepaspectratio, width=0.99\columnwidth]{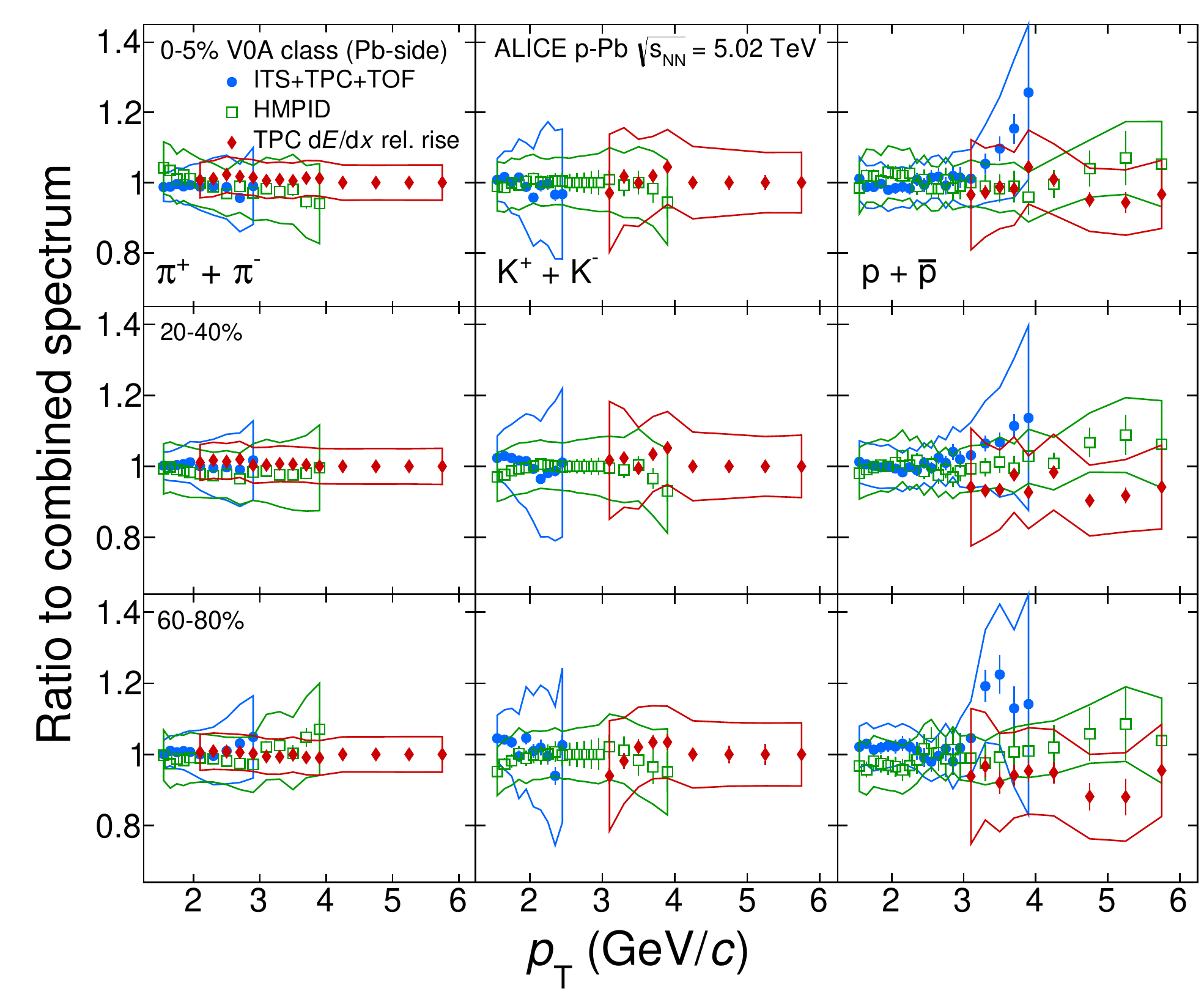}
  \caption{(Color online.) The ratio of individual spectra to the combined spectrum as a function of $p_{\rm T}$ for pions (left), kaons (middle), and protons (right). From top-to-bottom the rows show the V0A multiplicity class 0--5\%, 20--40\% and 60--80\%. Statistical and uncorrelated systematic uncertainties are shown as vertical error bars and error bands, respectively. Only the $p_{\rm T}$ ranges where individual analysis overlap are shown. See the text for further details.}
  \label{fig1:sec21}
\end{figure}

\section{Results and discussions}
\label{sec:results}

The total systematic uncertainty for all the spectra for a given particle species is factorized for each \pt interval into a multiplicity independent and multiplicity dependent systematic uncertainty. The transverse momentum distributions obtained from the different analyses are combined in the overlapping \pt region using a weighted average.  The weight for the combinations was done according to the total
systematic uncertainty to obtain the best overall precision. Since the systematic uncertainties due to normalization and tracking are common to all the analyses, they were added directly to the final combined results. The statistical uncertainties are much smaller and therefore neglected in the combination  weights. The multiplicity dependent systematic uncertainty for the combined spectra is also propagated using the same weights. For the results shown in this paper the full systematic uncertainty is always used, but the multiplicity correlated and uncorrelated systematic uncertainties are made available at HepData.   Figure~\ref{fig1:sec21} shows examples of the comparisons among the individual analyses and the combined \pt spectra, focusing on the overlapping \pt region. Within systematic and statistical uncertainties the new high-\pt results, measured with HMPID and TPC, agree with the published results~\cite{Abelev:2013haa}. Similar agreement is obtained for the \pt spectra in INEL \pp collisions at 7 TeV~\cite{Adam:2015qaa}.

\subsection{Transverse momentum spectra and nuclear modification factor}

\begin{figure}[th!f]
  \centering
  \includegraphics[keepaspectratio, width=0.9\columnwidth]{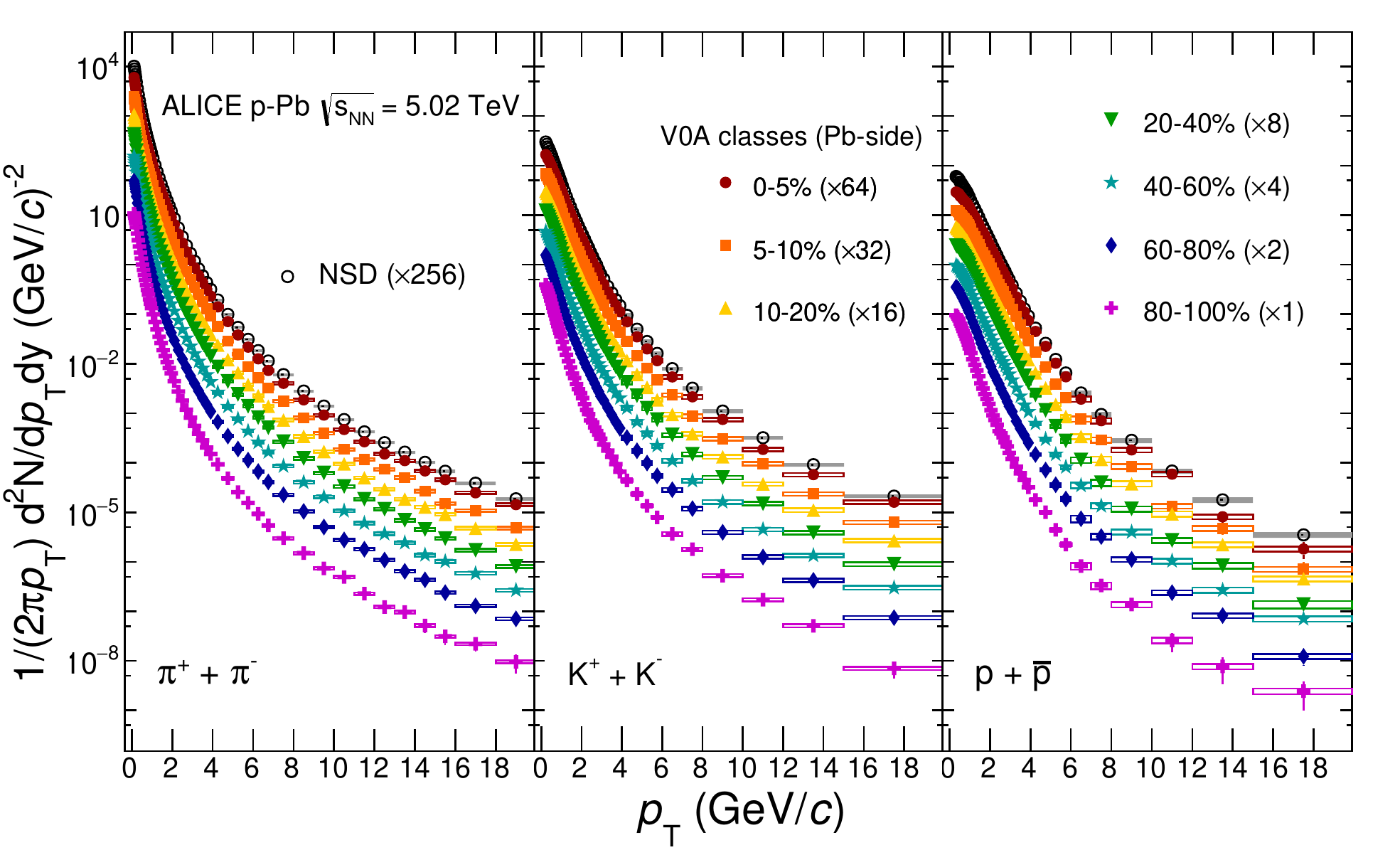}
  \caption{(Color online.) Transverse momentum spectra of charged pions (left), kaons (middle), and (anti)protons (right) measured in p--Pb collisions at $\sqrt{s_{\rm NN}}\,=5.02$\,TeV. Statistical and systematic uncertainties are plotted as vertical error bars and boxes, respectively. The spectra (measured for NSD events and for different V0A multiplicity classes) have been scaled by the indicated factors in the legend for better visibility.}
  \label{fig1:sec22}
\end{figure}

The combined charged pion, kaon and (anti)proton \pt spectra in \ppb collisions for different V0A multiplicity classes are shown in Fig.~\ref{fig1:sec22}.  As reported in~\cite{Abelev:2013haa}, for \pt
below 2--\gevc{3} the spectra behave like in \pbpb collisions, i.e., the \pt distributions become harder as the multiplicity increases and the change is most pronounced for protons and lambdas. In heavy-ion collisions this effect is commonly attributed to radial flow. For larger momenta, the spectra follow a power-law shape as expected from perturbative QCD.  

\begin{figure}[htb!f]
  \centering
  \includegraphics[keepaspectratio, width=0.9\columnwidth]{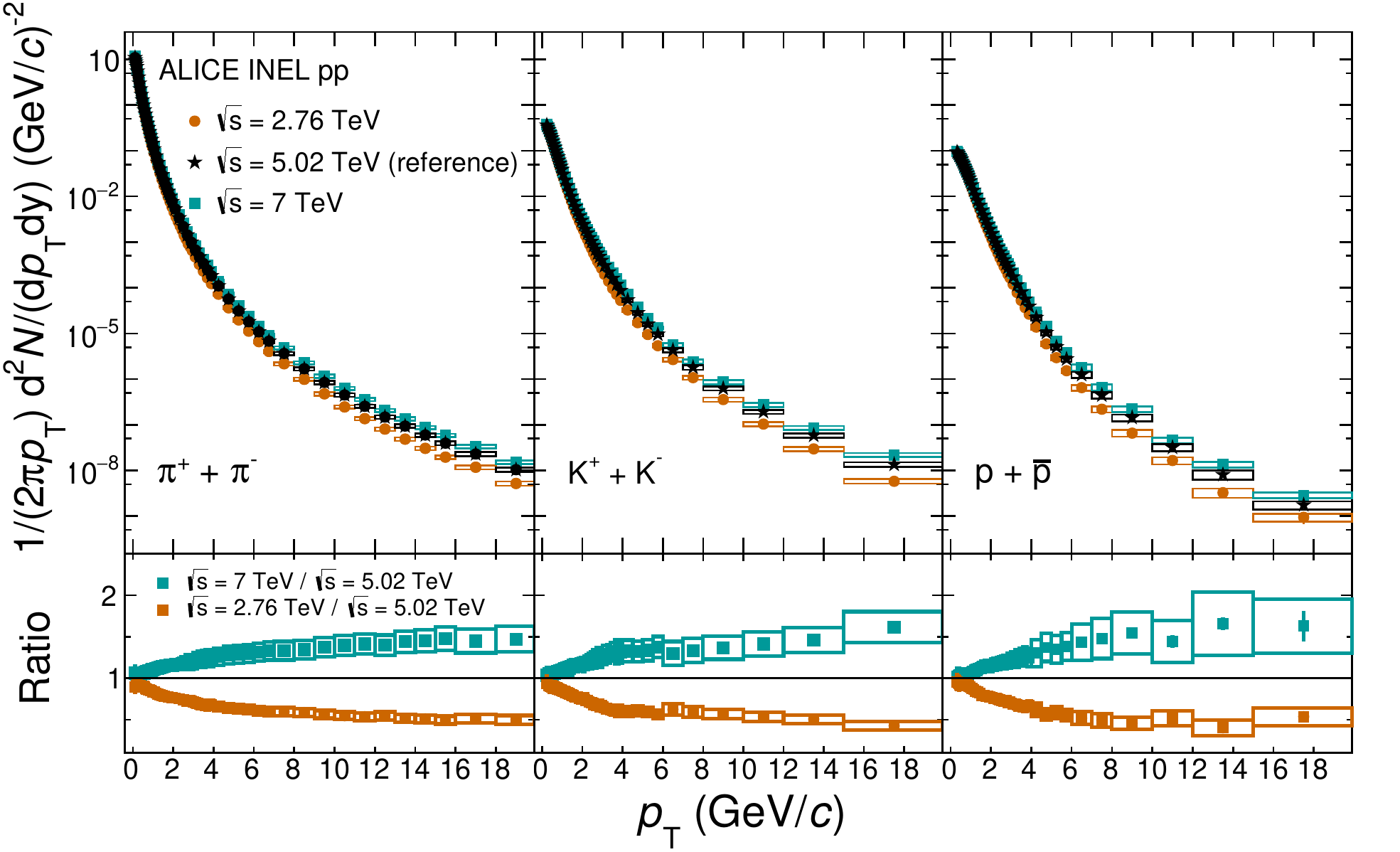}
  \caption{(Color online.) Transverse momentum spectra of charged pions (left), kaons (middle), and (anti)protons (right) measured in INEL pp collisions at $\sqrt{s}\,=2.76$ TeV and at $\sqrt{s}\,=7$ TeV. Statistical and systematic uncertainties are plotted as vertical error bars and boxes, respectively. The spectrum at $\sqrt{s}\,=5.02$\,TeV represents the reference in INEL pp collisions, constructed from measured spectra at $\sqrt{s}\,=2.76$ TeV and at $\sqrt{s}\,=7$ TeV. See the text for further details. Panels on the bottom show the ratio of the measured yields to the interpolated spectra. Only uncertainties of the interpolated spectra are shown.}
  \label{fig1:sec24}
\end{figure}

In order to quantify any particle species dependence of the nuclear effects in \ppb collisions, comparisons to reference \pt spectra in \pp collisions are needed. In the absence of \pp data at \sppt{5.02}, the reference spectra are
obtained by interpolating data measured at $\sppt{2.76}$ and at $\sppt{7}$. The invariant cross section for identified hadron production in INEL
\pp collisions, $1/(2\pi\pt)$ ${\rm d^{2}}\sigma^{\rm INEL}_{\rm pp}/{\rm d}y{\rm d}\pt$, is
interpolated in each \pt interval, assuming a power law dependence as a function of
$\sqrt{s}$. The method was cross-checked using events simulated by Pythia 8.201~\cite{Sjostrand:2014zea}, where the difference between the interpolated and the simulated reference was found to be negligible. The maximum relative systematic uncertainty of the spectra at \sppt{2.76} and at \sppt{7}  has been assigned as a systematic uncertainty to the reference.   In the transverse momentum interval $3<\pt<10$\,GeV/$c$, the total systematic uncertainties for pions and kaons are below 8.6\% and 10\%, respectively. While for (anti)protons it is 7.7\% and 18\% at 3\,GeV/$c$ and 10\,GeV/$c$, respectively. The invariant yields are shown in
Fig.~\ref{fig1:sec24}, where the interpolated \pt spectra are compared to
those measured in INEL \pp collisions at 2.76 TeV and 7
TeV.

The nuclear modification factor is then constructed as:
\begin{equation}
\Rppb = \frac{\text{d}^{2}N_{\text{pPb}}/\text{d}y\text{d}p_{\rm T}}{\left<T_{\rm pPb}\right>\text{d}^{2}\sigma^{\rm INEL}_{\rm pp}/\text{d}y\text{d}p_{\rm T}}
\end{equation}
where, for minimum bias (NSD) \ppb collisions the average nuclear overlap function, $\left<T_{\rm pPb}\right>$, is  $0.0983\pm0.0035 \text{\,mb}^{-1}$~\cite{ALICE:2012mj}. In absence of nuclear effects the \Rppb is expected to be one.  

\begin{figure}[htbp]
  \centering
  \includegraphics[keepaspectratio, width=0.6\columnwidth]{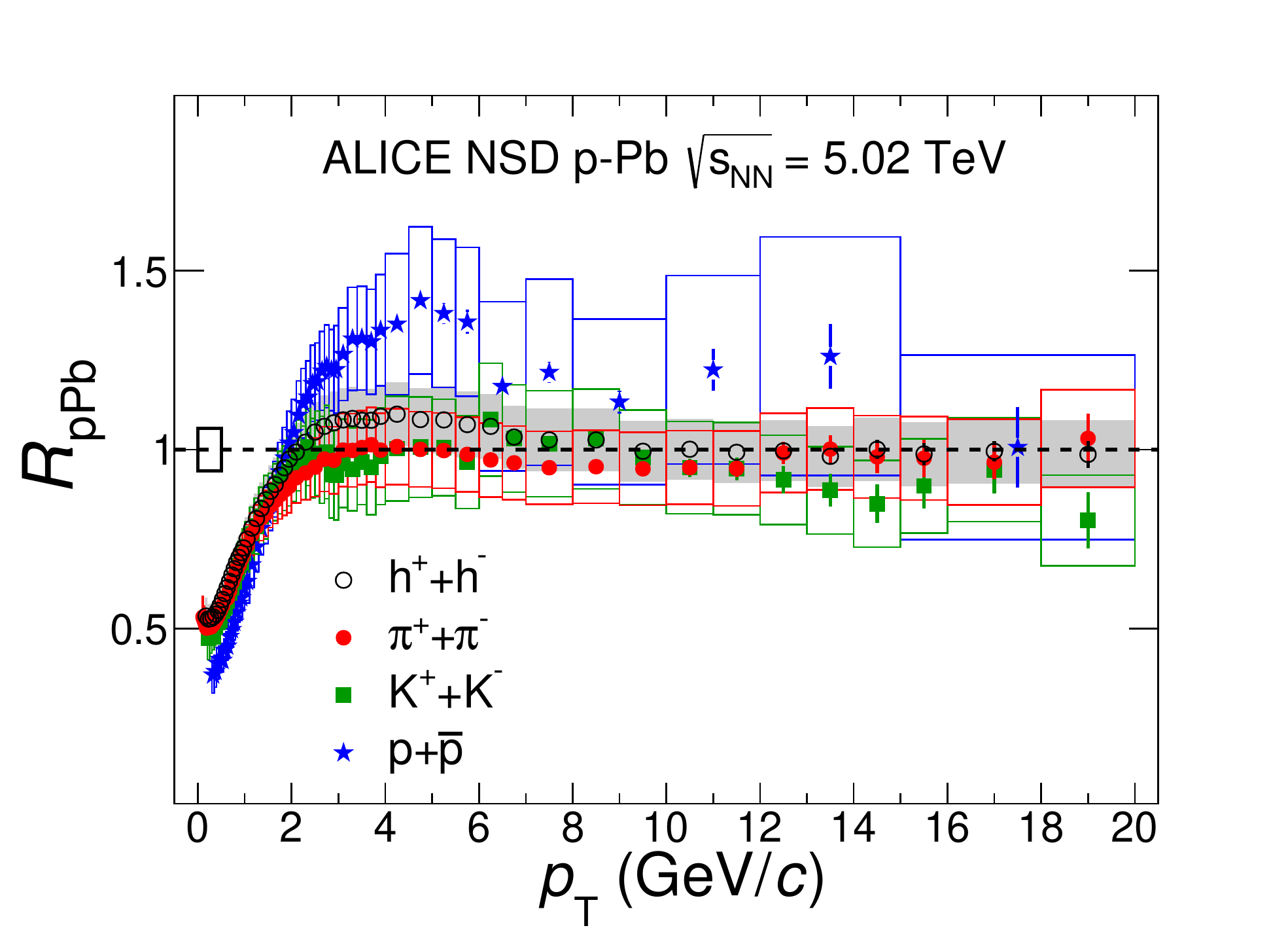}
  \caption{(Color online.) The nuclear modification factor $R_{\rm pPb}$ as a function of transverse momentum $p_{\rm T}$ for different particle species. The statistical and systematic uncertainties are shown as vertical error bars and boxes, respectively. The total normalization uncertainty is indicated by a vertical scale of the empty box at $p_{\rm T}=\text{0\,GeV/}c$ and $R_{\rm pPb}=\text{1}$. The result for inclusive charged hadrons~\cite{Abelev:2014dsa} is also shown.}
  \label{fig1:sec25}
\end{figure}

Figure~\ref{fig1:sec25} shows the identified hadron \Rppb compared to that for inclusive charged particles (h$^{\pm}$)~\cite{Abelev:2014dsa} in NSD \ppb events. At high \pt ($>\gevc{10}$), all nuclear modification factors are consistent with unity within systematic and statistical uncertainties.  Around \gevc{4}, where a prominent Cronin enhancement has been
seen at lower energies~\cite{PhysRevD.11.3105,PhysRevLett.31.1426}, the unidentified
charged hadron \Rppb is above unity, albeit  barely significant within systematic
uncertainties~\cite{Abelev:2014dsa}. Remarkably, the (anti)proton enhancement is $\sim$3 times larger than that for charged particles, while for charged pions and kaons the enhancement is below that of charged particles. The STAR and PHENIX Collaborations have observed a similar pattern at RHIC, where the nuclear modification factor for MB d-Au collisions, $R_{\rm dAu}$, in the range $2<\pt<5$\,GeV/$c$, is $1.24\pm0.13$ and $1.49\pm0.17$ for charged pions and (anti)protons, respectively~\cite{Adams:2006nd}. 

An enhancement of protons in the same \pt range is also observed in heavy-ion collisions~\cite{Abelev:2014laa,Adam:2015kca}, where it commonly is interpreted as radial-flow and has a strong centrality dependence. In the next section, we study the multiplicity dependence of the invariant yield ratios to see whether protons are more enhanced as a function of multiplicity than pions.

\subsection{Transverse momentum and multiplicity dependence of particle ratios}

\begin{figure}[htb!f]
  \centering
  \includegraphics[keepaspectratio, width=0.99\columnwidth]{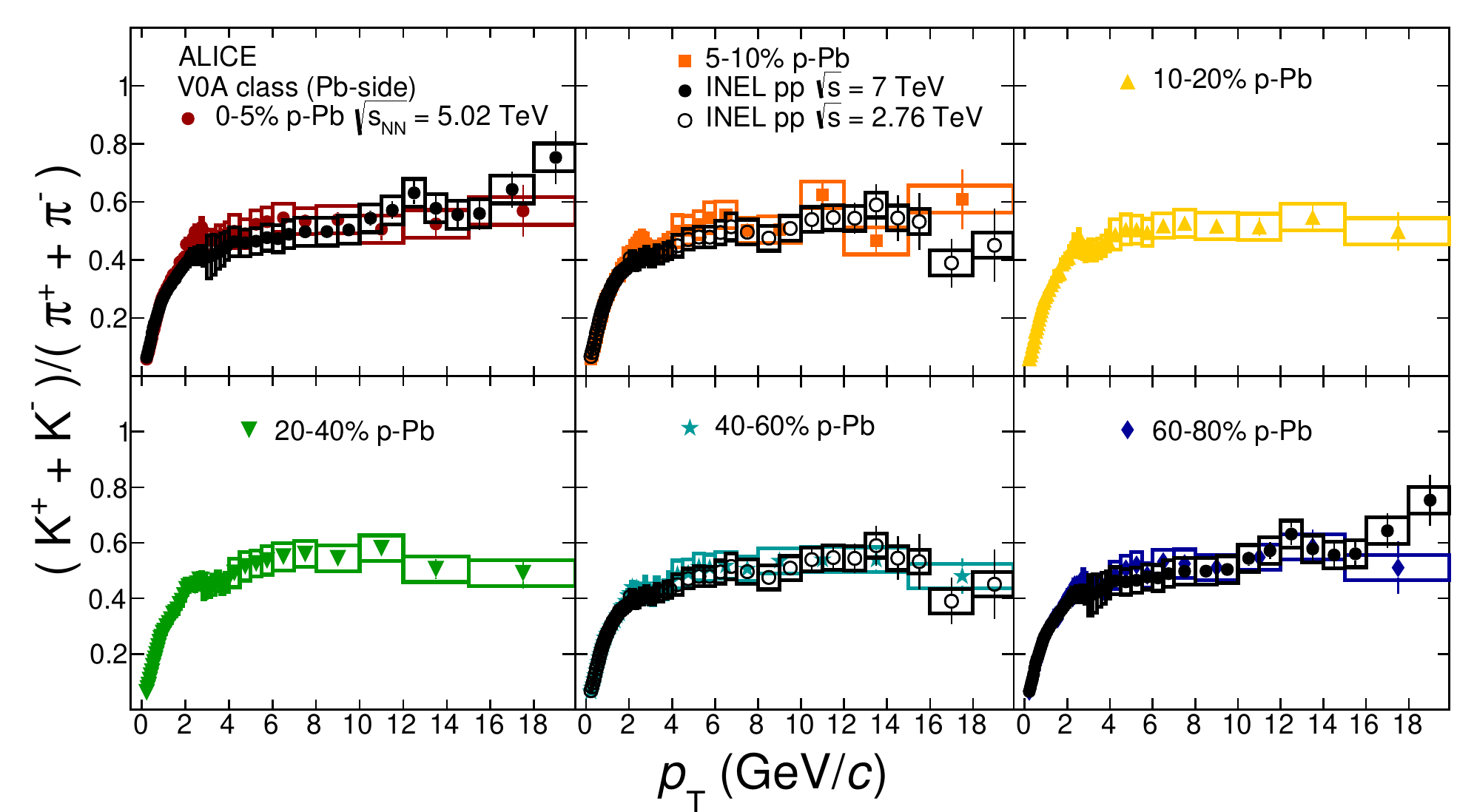}
  \includegraphics[keepaspectratio, width=0.99\columnwidth]{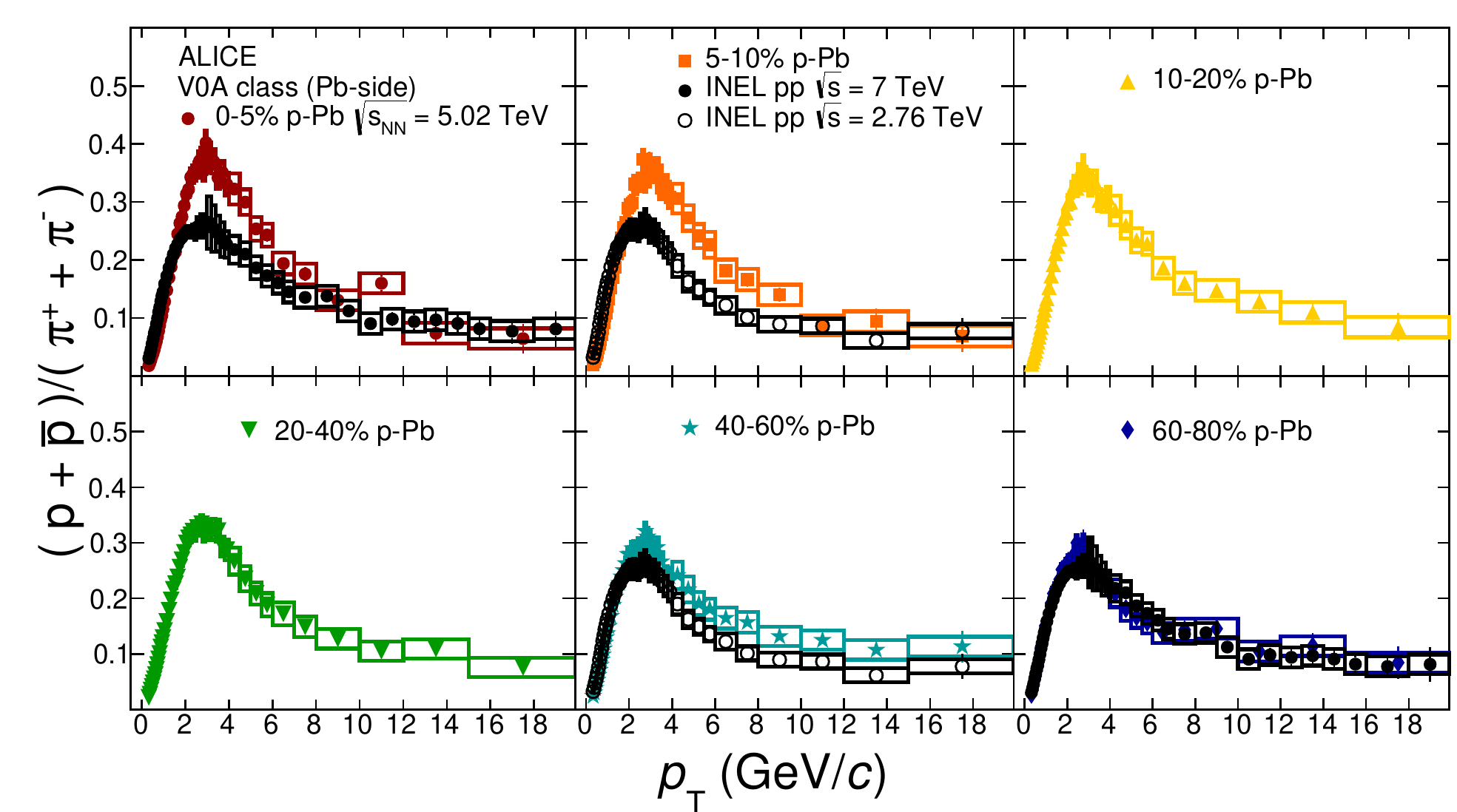}
  \caption{(Color online.) Kaon-to-pion (upper panel) and proton-to-pion (bottom panel) ratios as a function of $p_{\rm T}$ for different V0A multiplicity classes. Results for p--Pb collisions (full markers) are compared to the ratios measured in INEL pp collisions at 2.76 TeV~\cite{Abelev:2014laa} (empty circles) and at 7 TeV~\cite{Adam:2015qaa} (full circles). The statistical and systematic uncertainties are plotted as vertical error bars and boxes, respectively.}
  \label{fig1:sec23}
\end{figure}

The kaon-to-pion and the proton-to-pion ratios as a function of \pt for
different V0A multiplicity classes are shown in Fig.~\ref{fig1:sec23}. The
results for \ppb collisions are compared to those measured for INEL \pp
collisions at 2.76 TeV~\cite{Abelev:2014laa} and at 7 TeV~\cite{Adam:2015qaa}. Within systematic and statistical uncertainties, the \pt differential kaon-to-pion
ratios do not show any multiplicity dependence. In fact, the results are
similar to those for INEL \pp collisions at both energies. The \pt differential proton-to-pion ratios show a clear
multiplicity evolution at low and intermediate \pt ($<\gevc{10}$). This multiplicity evolution is qualitatively similar to the centrality
evolution observed in \pbpb collisions~\cite{Abelev:2014laa,Adam:2015kca}. 

\begin{figure}[tb!f]
  \centering
  \includegraphics[keepaspectratio, width=0.8\columnwidth]{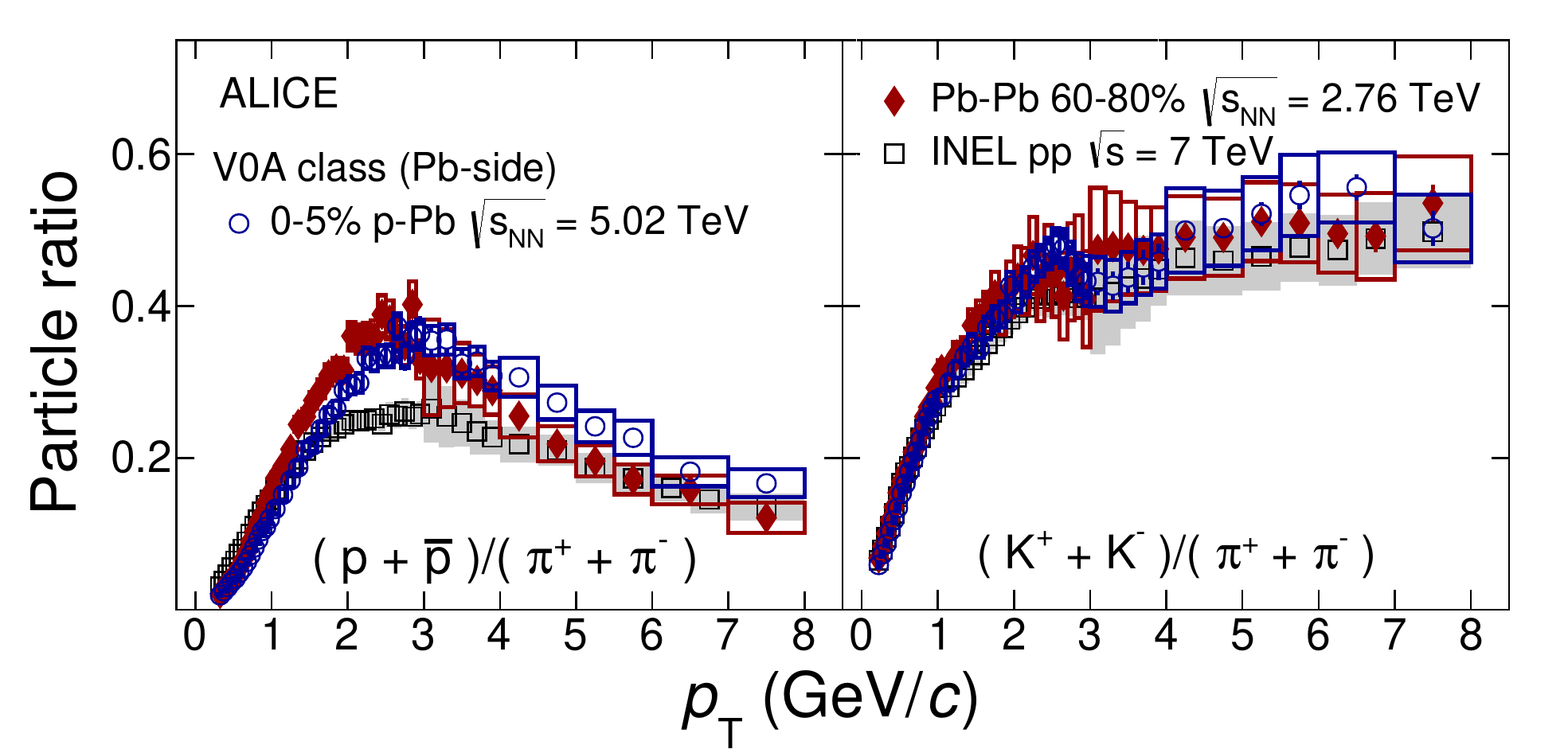}
  \caption{(Color online.) Particle ratios as a function of transverse momentum. Three different colliding systems are compared, pp (open squares), 0--5\% p--Pb (open circles) and 60--80\% Pb--Pb (full diamonds) collisions at $\sqrt{s_{\rm NN}}=$ 7 TeV, 5.02 TeV and 2.76 TeV, respectively. }
  \label{fig2:sec25}
\end{figure}

It is worth noting that the average multiplicities at mid-rapidity for peripheral \pbpb collisions (60--80\%) and high multiplicity \ppb collisions (0--5\% V0A multiplicity class) are very similar, $\langle {\rm d}N_{\rm ch}/{\rm d}\eta\rangle$$\sim$50. Even if the physical mechanisms for particle production could be different, it seems interesting to compare these systems with similar underlying activity as done in Fig.~\ref{fig2:sec25}, where INEL \sppt{7} \pp results are included as an approximate baseline. Within systematic and statistical uncertainties, the kaon-to-pion ratios are the same for all systems. On the other hand, the proton-to-pion ratios exhibit similar flow-like features for the \ppb and \pbpb systems, namely, the ratios are below the \pp baseline for $\pt<1$\,GeV/$c$ and above for $\pt>1.5$\,GeV/$c$. Quantitative differences are observed between \ppb and \pbpb results, but they can be attributed to the differences in the initial state overlap geometry and the beam energy. 

\begin{figure}[htb!F]
  \centering
  \includegraphics[keepaspectratio, width=0.8\columnwidth]{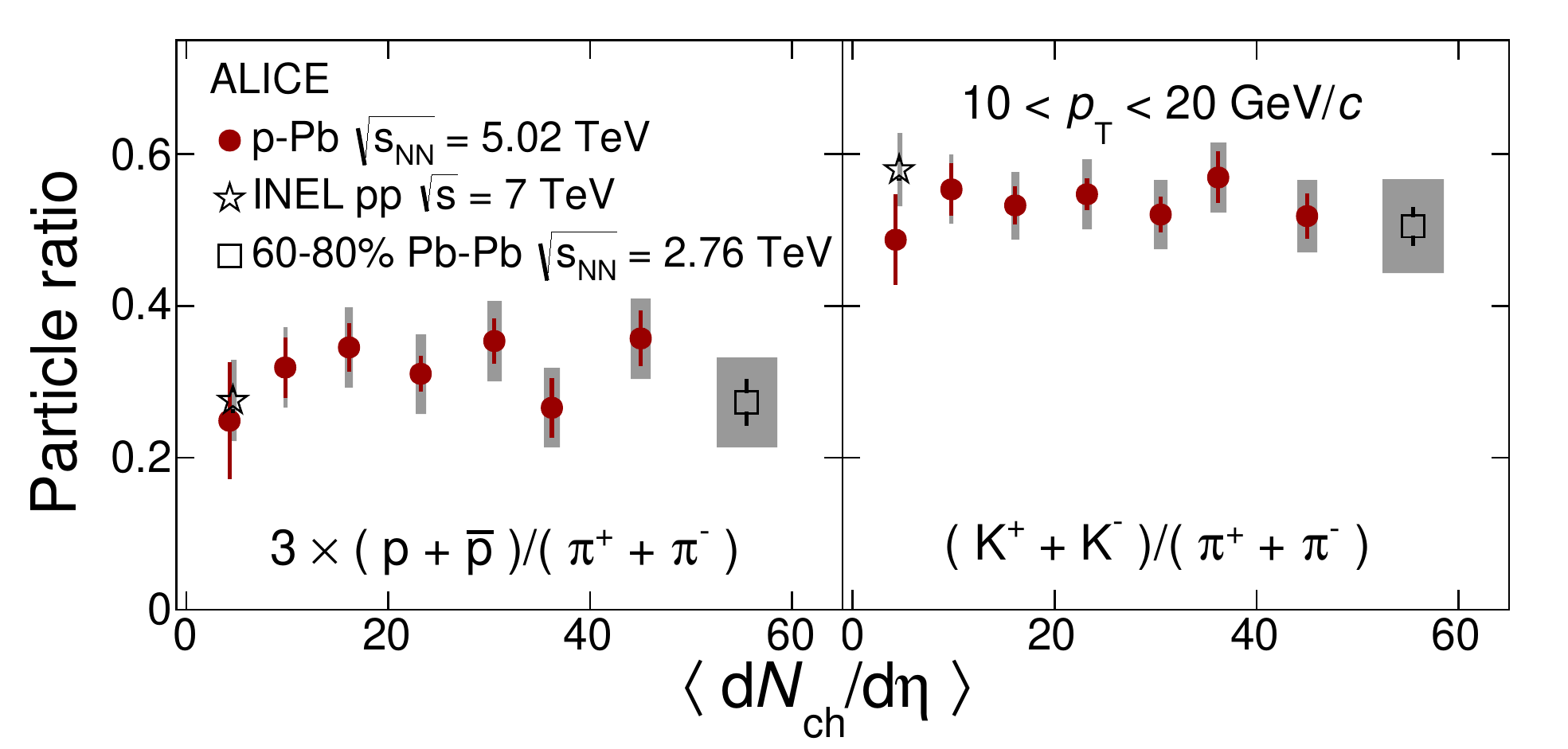}
  \caption{(Color online.) Particle ratios as a function of  $\langle {\rm d}N_{\rm ch}/{\rm d}\eta \rangle$ in each V0A multiplicity class (see~\cite{Abelev:2013haa} for more details). Three different colliding systems are compared: pp, p--Pb and peripheral Pb--Pb collisions. }
  \label{fig3:sec25}
\end{figure}

The results for the particle ratios suggest that the modification of the (anti)proton spectral shape going from \pp to \ppb collisions could play the dominant role in the Cronin enhancement observed for inclusive charged particle \Rppb at LHC energies. To confirm this picture one would have to study the nuclear modification factor as a function of multiplicity as we did in~\cite{Adam:2014qja}, where, the possible biases  in the evaluation of the multiplicity-dependent average nuclear overlap function $\left<T_{\rm pPb}\right>$ were discussed. These results will become available in the future.

In Fig.~\ref{fig3:sec25} we compare the particle ratios at high \pt ($10<\pt<20$\,GeV/$c$) measured in INEL  \sppt{7} \pp collisions, peripheral \pbpb collisions and the multiplicity dependent results in \ppb collisions.  Within statistical and systematic uncertainties, the ratios do not show any evolution with multiplicity. Moreover,  since it has been already reported that in \pbpb collisions they are centrality independent~\cite{Adam:2015kca} we conclude that they are system-size independent. 

The strong similarity of particle ratios as a function of multiplicity in \ppb and centrality in \pbpb collisions in the low,
intermediate, and high-\pt regions is striking. In general, the results  for
\ppb collisions appear to raise questions about the long standing ideas of
specific physics models for small and large systems~\cite{Antinori:2014xma}.  For example, in the low \pt publication~\cite{Abelev:2013haa},  hydrodynamic inspired fits gave higher
transverse expansion velocities ($\langle \beta_{\rm T} \rangle$) for \ppb than for \pbpb collisions. Hydrodynamics, which successfully describes many features of heavy-ion collisions, has been applied to small systems and can explain this effect~\cite{Shuryak:2013ke},  but care needs to be taken since its applicability to small systems is still under debate~\cite{Antinori:2014xma}. On the other hand, models like color
reconnection, where the soft and hard components are allowed to interact,
produce this kind of effects in \pp collisions~\cite{PhysRevLett.111.042001,Bierlich:2015rha}. Even more, the hard collisions which could be enhanced via the multiplicity selection in small systems, also contribute to increase $\langle \beta_{\rm T} \rangle$~\cite{Ortiz201578}. In general, color reconnection effects in \ppb and \pbpb collisions are under investigation and models for the effect of strong color fields in small systems are in general under development~\cite{Bierlich:2014xba}. Finally, it has been proposed that in d-Au collisions the recombination of soft and shower partons in the final state could explain the behavior of the nuclear modification factor at intermediate \pt~\cite{Hwa:2004zd}. The CMS Collaboration has found that the second-order ($v_2$) and the third-order ($v_3$) anisotropy harmonics
measured for ${\rm K}^0_{\rm S}$ and $\Lambda$ show constituent quark scaling in \ppb
collisions~\cite{Khachatryan:2014jra}.


\section{Conclusions}
\label{sec:conclusions}

We have reported on the charged pion, kaon and (anti)proton production up to
large transverse momenta ($\pt \leq \gevc{20}$) in \ppb collisions at $\sqrt{s_{\rm NN}}=$5.02\,TeV. The \pt spectra in \sppt{7} \pp collisions were also measured up
to \gevc{20} to allow the determination of the \sppt{5.02} \pp reference cross section using
the existing data at 2.76\,TeV and at 7\,TeV.

At intermediate \pt ($2 < \pt < \gevc{10}$), the (anti)proton \Rppb for non-single diffractive \ppb collisions was found to be significantly larger than those for pions and
kaons, in particular in the region where the Cronin peak was observed by
experiments at lower energies. Hence, the modest enhancement which we already reported for unidentified charged particles can  be attributed to the
modification of the proton spectral shape going from \pp to \ppb collisions. At high \pt the nuclear modification factors for charged pions, kaons and (anti)protons are consistent with unity within systematic and statistical uncertainties.

The enhancement of protons with respect to pions at intermediate \pt shows a
strong multiplicity dependence. This behavior is not observed for the
kaon-to-pion ratio. At high transverse momenta ($10 < \pt < \gevc{20}$) the
\pt integrated particle ratios are system-size independent for \pp, \ppb and \pbpb collisions. For a similar multiplicity at mid-rapidity, the \pt-differential particle ratios are alike for \ppb and \pbpb collisions over the broad \pt range reported in this paper.

\newenvironment{acknowledgement}{\relax}{\relax}
\begin{acknowledgement}
\section*{Acknowledgements}

The ALICE Collaboration would like to thank all its engineers and technicians for their invaluable contributions to the construction of the experiment and the CERN accelerator teams for the outstanding performance of the LHC complex.
The ALICE Collaboration gratefully acknowledges the resources and support provided by all Grid centres and the Worldwide LHC Computing Grid (WLCG) collaboration.
The ALICE Collaboration acknowledges the following funding agencies for their support in building and
running the ALICE detector:
State Committee of Science,  World Federation of Scientists (WFS)
and Swiss Fonds Kidagan, Armenia;
Conselho Nacional de Desenvolvimento Cient\'{\i}fico e Tecnol\'{o}gico (CNPq), Financiadora de Estudos e Projetos (FINEP),
Funda\c{c}\~{a}o de Amparo \`{a} Pesquisa do Estado de S\~{a}o Paulo (FAPESP);
National Natural Science Foundation of China (NSFC), the Chinese Ministry of Education (CMOE)
and the Ministry of Science and Technology of China (MSTC);
Ministry of Education and Youth of the Czech Republic;
Danish Natural Science Research Council, the Carlsberg Foundation and the Danish National Research Foundation;
The European Research Council under the European Community's Seventh Framework Programme;
Helsinki Institute of Physics and the Academy of Finland;
French CNRS-IN2P3, the `Region Pays de Loire', `Region Alsace', `Region Auvergne' and CEA, France;
German Bundesministerium fur Bildung, Wissenschaft, Forschung und Technologie (BMBF) and the Helmholtz Association;
General Secretariat for Research and Technology, Ministry of Development, Greece;
National Research, Development and Innovation Office (NKFIH), Hungary;
Department of Atomic Energy and Department of Science and Technology of the Government of India;
Istituto Nazionale di Fisica Nucleare (INFN) and Centro Fermi -
Museo Storico della Fisica e Centro Studi e Ricerche ``Enrico Fermi'', Italy;
Japan Society for the Promotion of Science (JSPS) KAKENHI and MEXT, Japan;
Joint Institute for Nuclear Research, Dubna;
National Research Foundation of Korea (NRF);
Consejo Nacional de Cienca y Tecnologia (CONACYT), Direccion General de Asuntos del Personal Academico(DGAPA), M\'{e}xico, Amerique Latine Formation academique - 
European Commission~(ALFA-EC) and the EPLANET Program~(European Particle Physics Latin American Network);
Stichting voor Fundamenteel Onderzoek der Materie (FOM) and the Nederlandse Organisatie voor Wetenschappelijk Onderzoek (NWO), Netherlands;
Research Council of Norway (NFR);
National Science Centre, Poland;
Ministry of National Education/Institute for Atomic Physics and National Council of Scientific Research in Higher Education~(CNCSI-UEFISCDI), Romania;
Ministry of Education and Science of Russian Federation, Russian
Academy of Sciences, Russian Federal Agency of Atomic Energy,
Russian Federal Agency for Science and Innovations and The Russian
Foundation for Basic Research;
Ministry of Education of Slovakia;
Department of Science and Technology, South Africa;
Centro de Investigaciones Energeticas, Medioambientales y Tecnologicas (CIEMAT), E-Infrastructure shared between Europe and Latin America (EELA), 
Ministerio de Econom\'{i}a y Competitividad (MINECO) of Spain, Xunta de Galicia (Conseller\'{\i}a de Educaci\'{o}n),
Centro de Aplicaciones Tecnológicas y Desarrollo Nuclear (CEA\-DEN), Cubaenerg\'{\i}a, Cuba, and IAEA (International Atomic Energy Agency);
Swedish Research Council (VR) and Knut $\&$ Alice Wallenberg
Foundation (KAW);
Ukraine Ministry of Education and Science;
United Kingdom Science and Technology Facilities Council (STFC);
The United States Department of Energy, the United States National
Science Foundation, the State of Texas, and the State of Ohio;
Ministry of Science, Education and Sports of Croatia and  Unity through Knowledge Fund, Croatia;
Council of Scientific and Industrial Research (CSIR), New Delhi, India;
Pontificia Universidad Cat\'{o}lica del Per\'{u}.
\end{acknowledgement}

\bibliographystyle{utphys}   
\bibliography{biblio}

\newpage
\appendix
\section{The ALICE Collaboration}
\label{app:collab}



\begingroup
\small
\begin{flushleft}
J.~Adam\Irefn{org40}\And
D.~Adamov\'{a}\Irefn{org84}\And
M.M.~Aggarwal\Irefn{org88}\And
G.~Aglieri Rinella\Irefn{org36}\And
M.~Agnello\Irefn{org110}\And
N.~Agrawal\Irefn{org48}\And
Z.~Ahammed\Irefn{org132}\And
S.~Ahmad\Irefn{org19}\And
S.U.~Ahn\Irefn{org68}\And
S.~Aiola\Irefn{org136}
% ...
\renewcommand\labelenumi{\textsuperscript{\theenumi}~}

\section*{Affiliation notes}
\renewcommand\theenumi{\roman{enumi}}
\begin{Authlist}
\item \Adef{0}Deceased
\item \Adef{idp1758928}{Also at: Georgia State University, Atlanta, Georgia, United States}
\item \Adef{idp3119664}{Also at: Also at Department of Applied Physics, Aligarh Muslim University, Aligarh, India}
\item \Adef{idp3827184}{Also at: M.V. Lomonosov Moscow State University, D.V. Skobeltsyn Institute of Nuclear, Physics, Moscow, Russia}
\end{Authlist}

\section*{Collaboration Institutes}
\renewcommand\theenumi{\arabic{enumi}~}
\begin{Authlist}

\item \Idef{org1}A.I. Alikhanyan National Science Laboratory (Yerevan Physics Institute) Foundation, Yerevan, Armenia
\item \Idef{org2}Benem\'{e}rita Universidad Aut\'{o}noma de Puebla, Puebla, Mexico
\item \Idef{org3}Bogolyubov Institute for Theoretical Physics, Kiev, Ukraine
\item \Idef{org4}Bose Institute, Department of Physics and Centre for Astroparticle Physics and Space Science (CAPSS), Kolkata, India
\item \Idef{org5}Budker Institute for Nuclear Physics, Novosibirsk, Russia
\item \Idef{org6}California Polytechnic State University, San Luis Obispo, California, United States
\item \Idef{org7}Central China Normal University, Wuhan, China
\item \Idef{org8}Centre de Calcul de l'IN2P3, Villeurbanne, France
\item \Idef{org9}Centro de Aplicaciones Tecnol\'{o}gicas y Desarrollo Nuclear (CEADEN), Havana, Cuba
\item \Idef{org10}Centro de Investigaciones Energ\'{e}ticas Medioambientales y Tecnol\'{o}gicas (CIEMAT), Madrid, Spain
\item \Idef{org11}Centro de Investigaci\'{o}n y de Estudios Avanzados (CINVESTAV), Mexico City and M\'{e}rida, Mexico
\item \Idef{org12}Centro Fermi - Museo Storico della Fisica e Centro Studi e Ricerche ``Enrico Fermi'', Rome, Italy
\item \Idef{org13}Chicago State University, Chicago, Illinois, USA
\item \Idef{org14}China Institute of Atomic Energy, Beijing, China
\item \Idef{org15}Commissariat \`{a} l'Energie Atomique, IRFU, Saclay, France
\item \Idef{org16}COMSATS Institute of Information Technology (CIIT), Islamabad, Pakistan
\item \Idef{org17}Departamento de F\'{\i}sica de Part\'{\i}culas and IGFAE, Universidad de Santiago de Compostela, Santiago de Compostela, Spain
\item \Idef{org18}Department of Physics and Technology, University of Bergen, Bergen, Norway
\item \Idef{org19}Department of Physics, Aligarh Muslim University, Aligarh, India
\item \Idef{org20}Department of Physics, Ohio State University, Columbus, Ohio, United States
\item \Idef{org21}Department of Physics, Sejong University, Seoul, South Korea
\item \Idef{org22}Department of Physics, University of Oslo, Oslo, Norway
\item \Idef{org23}Dipartimento di Elettrotecnica ed Elettronica del Politecnico, Bari, Italy
\item \Idef{org24}Dipartimento di Fisica dell'Universit\`{a} 'La Sapienza' and Sezione INFN Rome, Italy
\item \Idef{org25}Dipartimento di Fisica dell'Universit\`{a} and Sezione INFN, Cagliari, Italy
\item \Idef{org26}Dipartimento di Fisica dell'Universit\`{a} and Sezione INFN, Trieste, Italy
\item \Idef{org27}Dipartimento di Fisica dell'Universit\`{a} and Sezione INFN, Turin, Italy
\item \Idef{org28}Dipartimento di Fisica e Astronomia dell'Universit\`{a} and Sezione INFN, Bologna, Italy
\item \Idef{org29}Dipartimento di Fisica e Astronomia dell'Universit\`{a} and Sezione INFN, Catania, Italy
\item \Idef{org30}Dipartimento di Fisica e Astronomia dell'Universit\`{a} and Sezione INFN, Padova, Italy
\item \Idef{org31}Dipartimento di Fisica `E.R.~Caianiello' dell'Universit\`{a} and Gruppo Collegato INFN, Salerno, Italy
\item \Idef{org32}Dipartimento di Scienze e Innovazione Tecnologica dell'Universit\`{a} del  Piemonte Orientale and Gruppo Collegato INFN, Alessandria, Italy
\item \Idef{org33}Dipartimento Interateneo di Fisica `M.~Merlin' and Sezione INFN, Bari, Italy
\item \Idef{org34}Division of Experimental High Energy Physics, University of Lund, Lund, Sweden
\item \Idef{org35}Eberhard Karls Universit\"{a}t T\"{u}bingen, T\"{u}bingen, Germany
\item \Idef{org36}European Organization for Nuclear Research (CERN), Geneva, Switzerland
\item \Idef{org37}Excellence Cluster Universe, Technische Universit\"{a}t M\"{u}nchen, Munich, Germany
\item \Idef{org38}Faculty of Engineering, Bergen University College, Bergen, Norway
\item \Idef{org39}Faculty of Mathematics, Physics and Informatics, Comenius University, Bratislava, Slovakia
\item \Idef{org40}Faculty of Nuclear Sciences and Physical Engineering, Czech Technical University in Prague, Prague, Czech Republic
\item \Idef{org41}Faculty of Science, P.J.~\v{S}af\'{a}rik University, Ko\v{s}ice, Slovakia
\item \Idef{org42}Faculty of Technology, Buskerud and Vestfold University College, Vestfold, Norway
\item \Idef{org43}Frankfurt Institute for Advanced Studies, Johann Wolfgang Goethe-Universit\"{a}t Frankfurt, Frankfurt, Germany
\item \Idef{org44}Gangneung-Wonju National University, Gangneung, South Korea
\item \Idef{org45}Gauhati University, Department of Physics, Guwahati, India
\item \Idef{org46}Helsinki Institute of Physics (HIP), Helsinki, Finland
\item \Idef{org47}Hiroshima University, Hiroshima, Japan
\item \Idef{org48}Indian Institute of Technology Bombay (IIT), Mumbai, India
\item \Idef{org49}Indian Institute of Technology Indore, Indore (IITI), India
\item \Idef{org50}Inha University, Incheon, South Korea
\item \Idef{org51}Institut de Physique Nucl\'eaire d'Orsay (IPNO), Universit\'e Paris-Sud, CNRS-IN2P3, Orsay, France
\item \Idef{org52}Institut f\"{u}r Informatik, Johann Wolfgang Goethe-Universit\"{a}t Frankfurt, Frankfurt, Germany
\item \Idef{org53}Institut f\"{u}r Kernphysik, Johann Wolfgang Goethe-Universit\"{a}t Frankfurt, Frankfurt, Germany
\item \Idef{org54}Institut f\"{u}r Kernphysik, Westf\"{a}lische Wilhelms-Universit\"{a}t M\"{u}nster, M\"{u}nster, Germany
\item \Idef{org55}Institut Pluridisciplinaire Hubert Curien (IPHC), Universit\'{e} de Strasbourg, CNRS-IN2P3, Strasbourg, France
\item \Idef{org56}Institute for Nuclear Research, Academy of Sciences, Moscow, Russia
\item \Idef{org57}Institute for Subatomic Physics of Utrecht University, Utrecht, Netherlands
\item \Idef{org58}Institute for Theoretical and Experimental Physics, Moscow, Russia
\item \Idef{org59}Institute of Experimental Physics, Slovak Academy of Sciences, Ko\v{s}ice, Slovakia
\item \Idef{org60}Institute of Physics, Academy of Sciences of the Czech Republic, Prague, Czech Republic
\item \Idef{org61}Institute of Physics, Bhubaneswar, India
\item \Idef{org62}Institute of Space Science (ISS), Bucharest, Romania
\item \Idef{org63}Instituto de Ciencias Nucleares, Universidad Nacional Aut\'{o}noma de M\'{e}xico, Mexico City, Mexico
\item \Idef{org64}Instituto de F\'{\i}sica, Universidad Nacional Aut\'{o}noma de M\'{e}xico, Mexico City, Mexico
\item \Idef{org65}iThemba LABS, National Research Foundation, Somerset West, South Africa
\item \Idef{org66}Joint Institute for Nuclear Research (JINR), Dubna, Russia
\item \Idef{org67}Konkuk University, Seoul, South Korea
\item \Idef{org68}Korea Institute of Science and Technology Information, Daejeon, South Korea
\item \Idef{org69}KTO Karatay University, Konya, Turkey
\item \Idef{org70}Laboratoire de Physique Corpusculaire (LPC), Clermont Universit\'{e}, Universit\'{e} Blaise Pascal, CNRS--IN2P3, Clermont-Ferrand, France
\item \Idef{org71}Laboratoire de Physique Subatomique et de Cosmologie, Universit\'{e} Grenoble-Alpes, CNRS-IN2P3, Grenoble, France
\item \Idef{org72}Laboratori Nazionali di Frascati, INFN, Frascati, Italy
\item \Idef{org73}Laboratori Nazionali di Legnaro, INFN, Legnaro, Italy
\item \Idef{org74}Lawrence Berkeley National Laboratory, Berkeley, California, United States
\item \Idef{org75}Moscow Engineering Physics Institute, Moscow, Russia
\item \Idef{org76}Nagasaki Institute of Applied Science, Nagasaki, Japan
\item \Idef{org77}National Centre for Nuclear Studies, Warsaw, Poland
\item \Idef{org78}National Institute for Physics and Nuclear Engineering, Bucharest, Romania
\item \Idef{org79}National Institute of Science Education and Research, Bhubaneswar, India
\item \Idef{org80}National Research Centre Kurchatov Institute, Moscow, Russia
\item \Idef{org81}Niels Bohr Institute, University of Copenhagen, Copenhagen, Denmark
\item \Idef{org82}Nikhef, Nationaal instituut voor subatomaire fysica, Amsterdam, Netherlands
\item \Idef{org83}Nuclear Physics Group, STFC Daresbury Laboratory, Daresbury, United Kingdom
\item \Idef{org84}Nuclear Physics Institute, Academy of Sciences of the Czech Republic, \v{R}e\v{z} u Prahy, Czech Republic
\item \Idef{org85}Oak Ridge National Laboratory, Oak Ridge, Tennessee, United States
\item \Idef{org86}Petersburg Nuclear Physics Institute, Gatchina, Russia
\item \Idef{org87}Physics Department, Creighton University, Omaha, Nebraska, United States
\item \Idef{org88}Physics Department, Panjab University, Chandigarh, India
\item \Idef{org89}Physics Department, University of Athens, Athens, Greece
\item \Idef{org90}Physics Department, University of Cape Town, Cape Town, South Africa
\item \Idef{org91}Physics Department, University of Jammu, Jammu, India
\item \Idef{org92}Physics Department, University of Rajasthan, Jaipur, India
\item \Idef{org93}Physik Department, Technische Universit\"{a}t M\"{u}nchen, Munich, Germany
\item \Idef{org94}Physikalisches Institut, Ruprecht-Karls-Universit\"{a}t Heidelberg, Heidelberg, Germany
\item \Idef{org95}Purdue University, West Lafayette, Indiana, United States
\item \Idef{org96}Pusan National University, Pusan, South Korea
\item \Idef{org97}Research Division and ExtreMe Matter Institute EMMI, GSI Helmholtzzentrum f\"ur Schwerionenforschung, Darmstadt, Germany
\item \Idef{org98}Rudjer Bo\v{s}kovi\'{c} Institute, Zagreb, Croatia
\item \Idef{org99}Russian Federal Nuclear Center (VNIIEF), Sarov, Russia
\item \Idef{org100}Saha Institute of Nuclear Physics, Kolkata, India
\item \Idef{org101}School of Physics and Astronomy, University of Birmingham, Birmingham, United Kingdom
\item \Idef{org102}Secci\'{o}n F\'{\i}sica, Departamento de Ciencias, Pontificia Universidad Cat\'{o}lica del Per\'{u}, Lima, Peru
\item \Idef{org103}Sezione INFN, Bari, Italy
\item \Idef{org104}Sezione INFN, Bologna, Italy
\item \Idef{org105}Sezione INFN, Cagliari, Italy
\item \Idef{org106}Sezione INFN, Catania, Italy
\item \Idef{org107}Sezione INFN, Padova, Italy
\item \Idef{org108}Sezione INFN, Rome, Italy
\item \Idef{org109}Sezione INFN, Trieste, Italy
\item \Idef{org110}Sezione INFN, Turin, Italy
\item \Idef{org111}SSC IHEP of NRC Kurchatov institute, Protvino, Russia
\item \Idef{org112}Stefan Meyer Institut f\"{u}r Subatomare Physik (SMI), Vienna, Austria
\item \Idef{org113}SUBATECH, Ecole des Mines de Nantes, Universit\'{e} de Nantes, CNRS-IN2P3, Nantes, France
\item \Idef{org114}Suranaree University of Technology, Nakhon Ratchasima, Thailand
\item \Idef{org115}Technical University of Ko\v{s}ice, Ko\v{s}ice, Slovakia
\item \Idef{org116}Technical University of Split FESB, Split, Croatia
\item \Idef{org117}The Henryk Niewodniczanski Institute of Nuclear Physics, Polish Academy of Sciences, Cracow, Poland
\item \Idef{org118}The University of Texas at Austin, Physics Department, Austin, Texas, USA
\item \Idef{org119}Universidad Aut\'{o}noma de Sinaloa, Culiac\'{a}n, Mexico
\item \Idef{org120}Universidade de S\~{a}o Paulo (USP), S\~{a}o Paulo, Brazil
\item \Idef{org121}Universidade Estadual de Campinas (UNICAMP), Campinas, Brazil
\item \Idef{org122}University of Houston, Houston, Texas, United States
\item \Idef{org123}University of Jyv\"{a}skyl\"{a}, Jyv\"{a}skyl\"{a}, Finland
\item \Idef{org124}University of Liverpool, Liverpool, United Kingdom
\item \Idef{org125}University of Tennessee, Knoxville, Tennessee, United States
\item \Idef{org126}University of the Witwatersrand, Johannesburg, South Africa
\item \Idef{org127}University of Tokyo, Tokyo, Japan
\item \Idef{org128}University of Tsukuba, Tsukuba, Japan
\item \Idef{org129}University of Zagreb, Zagreb, Croatia
\item \Idef{org130}Universit\'{e} de Lyon, Universit\'{e} Lyon 1, CNRS/IN2P3, IPN-Lyon, Villeurbanne, France
\item \Idef{org131}V.~Fock Institute for Physics, St. Petersburg State University, St. Petersburg, Russia
\item \Idef{org132}Variable Energy Cyclotron Centre, Kolkata, India
\item \Idef{org133}Warsaw University of Technology, Warsaw, Poland
\item \Idef{org134}Wayne State University, Detroit, Michigan, United States
\item \Idef{org135}Wigner Research Centre for Physics, Hungarian Academy of Sciences, Budapest, Hungary
\item \Idef{org136}Yale University, New Haven, Connecticut, United States
\item \Idef{org137}Yonsei University, Seoul, South Korea
\item \Idef{org138}Zentrum f\"{u}r Technologietransfer und Telekommunikation (ZTT), Fachhochschule Worms, Worms, Germany
\end{Authlist}
\endgroup

\end{document}